\documentclass[a4paper,11pt]{article} % default is 10pt
\pdfoutput=1

\usepackage{jheppub}
\usepackage{amsmath,amssymb,xfrac,framed,verbatim,amsthm}
\usepackage{mathrsfs,accents,bbm,bbold} % for mathematical symbols
\usepackage{slashed,mathtools,textcomp,setspace} % for Dirac slash
\usepackage{graphicx,enumerate,bm} % for \includegraphics command and options
\newcommand\pmat[1] {\begin{pmatrix}#1\end{pmatrix}}

\newcommand{\wt}[1]{\widetilde{#1}}
\newcommand{\tr}{\text{Tr}}
\newcommand{\str}{\text{tr}}

\newcommand{\mc}[1]{\mathcal{#1}}

\newcommand{\ol}[1]{\overline{#1}}
\newcommand{\ud}{\text{d}}

\newcommand{\tl}{\tilde}

\renewcommand{\i}{\text{i}}
\newcommand{\e}{\text{e}}
\newcommand{\sgn}{\text{sgn}}

\newcommand{\mg}{\bm{g}}
\newcommand{\mM}{\bm{M}}
\newcommand{\bx}{\bm{x}}
\newcommand{\bp}{\bm{p}}
\newcommand{\mS}{\bm{S}}
\newcommand{\mW}{\bm{W}}

\newcommand{\cA}{\mathcal{A}}
\newcommand{\cD}{\mathcal{D}}
\newcommand{\cF}{\mathcal{F}}


\title{{\bf Meson spectrum of $\text{SU}(2)$ QCD$_{1+1}$ with \\
    Quarks in Large Representations}}

\author[a,1]{Anurag Kaushal,\note{anuragkaushal314@gmail.com}}
\author[a,2]{Naveen S.~Prabhakar,\note{naveen.s.prabhakar@gmail.com}}
\author[a,3]{Spenta R.~Wadia,\note{spenta.wadia@icts.res.in}}

\affiliation[a]{ International Centre for Theoretical Sciences-Tata Institute of Fundamental Research, Shivakote, Bengaluru 560089, India.}

\abstract{We consider $\text{SU}(2)$ quantum chromodynamics in $1+1$
dimensions with a single quark in the spin $J$ representation of the
gauge group and study the theory in the large $J$ limit where the
gauge coupling $g^2 \to 0$ and $J \to \infty$ with
$\lambda = g^2 J^2$ fixed. We work with a Dirac spinor field for
arbitrary $J$, and with a Majorana spinor for integer $J$ since the
integer spin representations of $\text{SU}(2)$ are real, and analyze
the two cases separately.

The theory is reformulated in terms of global colour non-singlet
fermion bilocal operators which satisfy a
$W_\infty \times \text{U}(2J+1)$ algebra. In the large $J$ limit,
the dynamics of the bilocal fields is captured by fluctuations along
a particular coadjoint orbit of the $W_\infty$ algebra. We show that
the global colour-singlet sector of the bilocal field fluctuations
satisfy the same integral equation for meson wavefunctions that
appears in the 't Hooft model. For Majorana spinors in the integer
spin $J$ representation, the Majorana condition projects out half of
the meson spectrum, as a result of which the linear spacing of the
asymptotic meson spectrum for Majorana fermions is double that of
Dirac fermions. The Majorana condition also projects out the zero
mass bound state that is present for the Dirac quark at zero quark
mass.

We also consider the formulation of
the model in terms of local charge densities and compute the quark
spectral function in the large $J$ limit: we see evidence for the
absence of a pole in the quark propagator.}

\begin{document}

\maketitle

\section{Introduction and summary}

Quantum chromodynamics in $1+1$ dimensions is expected to exhibit
confinement due to the Coulomb potential being linear in two spacetime
dimensions. The $\text{SU}(N)$ theory with quarks in the fundamental
representation was solved by 't Hooft in the planar large $N$ limit
\cite{tHooft:1974pnl,tHooft:1973alw}, and was further analyzed
extensively by many authors
\cite{Callan:1975ps,Einhorn:1976uz,Bars:1977ud}. Indeed, at leading
order in large $N$, it was first shown in \cite{tHooft:1974pnl} that
(1) there is no pole in the fermion-antifermion two-point function,
indicating that there is no asymptotic coloured state, and (2) the
gauge invariant quark-antiquark wavefunction satisfies an integral
eigenvalue equation whose solutions describe an infinite, discrete
spectrum of bound states.

The meson spectrum was derived in an entirely different way in
\cite{Dhar:1994ib} by noting that the Hamiltonian of the model in
lightcone gauge can be written in terms of the bilocal field
$\sum_i \psi^i(x)\bar\psi_{i}(y)$ which is nothing but the manifestly
gauge-invariant Wilson line in the lightcone gauge. It was shown in
\cite{Dhar:1994ib} that the bilocal field (i.e., the gauge invariant
Wilson line in lightcone gauge) satisfies a $W_\infty$ algebra. In the
large $N$ limit, the gauge-invariant Wilson line behaves
semiclassically due to factorization, and the classical phase space of
the bilocal field is an appropriate coadjoint orbit of the $W_\infty$
algebra \cite{Dhar:1994ib, Dhar:1993jc, Dhar:1992hr,Das:1991uta}. The
equations of motion for fluctuations along this coadjoint orbit then
directly gave the integral equation of 't Hooft for the gauge
invariant quark-antiquark wavefunction.

In this paper, we consider $\text{SU}(2)$ gauge theory in $1+1$
dimensions with a single quark in the spin $J$ representation of the
gauge group with the spin $J$ being very large.\footnote{In $3+1$
  dimensions, the theory is no longer asymptotically free in an
  analogous large $J$ limit. Indeed, the $\beta$-function for
  $\text{SU}(2)$ QCD with spin $J$ quarks is
  $\beta(g) = -\frac{g^3}{(4\pi)^2}\left(\frac{22}{3} - \frac{4}{3}
    \frac{J(J+1)(2J+1)}{3}\right)$, see, e.g.,
  \cite[Eq.~(16.134)]{Peskin:1995ev}. Clearly, the second term
  dominates over the first in the large $J$ limit, and flips the sign
  of the $\beta$ function.} The quark field will be a Majorana spinor
since integer $J$ representations of $\text{SU}(2)$ are real, whereas
we consider Dirac fermions for arbitrary $J$. We perform lightcone
quantization \emph{a la} 't Hooft \cite{tHooft:1974pnl} by choosing
the lightcone direction $x^+$ as time and work in the lightcone gauge
$A_- = 0$. The advantage of lightcone quantization is that the sole
dynamical degree of freedom is the chiral fermion $\psi_-(x)$. We
solve for the meson spectrum of the model in a 't Hooft-like large $J$
limit where $J \to \infty$ and the gauge coupling $g \to 0$ with
$\lambda = g^2 J^2$ kept fixed.\footnote{The theory with $J=1$
  corresponds to the $\text{SU}(2)$ theory with Majorana quarks in the
  adjoint representation whose spectrum has been numerically computed
  via discrete lightcone quantization (DLCQ) in
  \cite{Dempsey:2022uie}. See \cite{Narayanan:2023jmi} for a DLCQ
  analysis of the $\text{SU}(2)$ model with Majorana quarks in the
  $J=2,3,4$ representations.}

In Section \ref{mainsec}, we describe the large $J$ limit of the
theory. In taking the large $J$ limit, we treat the spin $J$
representation semiclassically and replace the discrete representation
index $m$ which runs over $-J,-J+1,\ldots,J-1,J$, by a continuum angle
$\theta \in [0,\pi]$ with $\cos\theta = m/J$. In this continuum
description, the fermion components $\psi^m(x)$ coalesce into a field
$\psi(\theta,x)$ along the $\theta$ direction, and we work with this
continuum field in our analysis. The four-fermi term in the
Hamiltonian takes a nice suggestive form where a potential also
appears in $\theta$ space, in addition to the Coulomb potential in
spacetime. Schematically, it takes the form
\begin{equation}\label{ffJ}
\int_{x^-,y^-}  \int_{\theta,\theta'} \psi^*_-(\theta,x)\psi_-(\theta,x) \cos(\theta - \theta') |x^- - y^-|  \psi^*_-(\theta',y) \psi_-(\theta',y)\ ,
\end{equation} 
where, the factor $|x^- - y^-|$ is the Coulomb potential along the
`spatial' direction $x^-$, and the integrals over $\theta$, $\theta'$
are the continuum versions of the sums over the representation indices
on the fermions.

Our subsequent analysis in Section \ref{thooftsec} is based on the
$W_\infty$ algebra and its coadjoint orbits that appear in
\cite{Dhar:1994ib}, but with one big conceptual difference. For the
$\text{SU}(N)$ theory with fundamental quarks considered in
\cite{Dhar:1994ib}, the effective four-fermi interaction due to the
Coulomb force between the quarks could be written as a product of two
bilocal fields which were in fact gauge invariant Wilson lines
evaluated in lightcone gauge. In our model, the four-fermi term cannot
be written in terms of such gauge invariant Wilson lines even in
lightcone gauge due to the potential in representation space (see
equation \eqref{ffJ}). However, the four-fermi term can be written in
terms of the \emph{non gauge-singlet} bilocal field
$\psi^i_-(x) \psi^*_{j-}(y)$ and hence we work with these
variables.\footnote{Such non colour-singlet variables have been used in
  other contexts, e.g., \cite{Mandal:2009vz}.}

The non gauge-singlet bilocal fields satisfy a
$W_\infty \times U(2J+1)$ algebra and lie on a particular coadjoint
orbit of $W_\infty \times U(2J+1)$. The equation of motion for
fluctuations along the coadjoint orbit gives an integral equation, but
now involving the non gauge-singlet bilocal field. Tracing over the
gauge indices of the non gauge-singlet bilocal field, we then get an
integral equation for the gauge invariant bilocal field
(gauge invariant because it is the gauge invariant Wilson line in
lightcone gauge) which is the same integral equation that was derived
by 't Hooft \cite{tHooft:1974pnl}. In
\cite{Bardeen:1975gx,Bars:1975dd,Bars:1976re}, the meson spectrum of
the large $N$ 't Hooft model was interpreted as the spectrum of a
string in two spacetime dimensions with massive particles at the ends
by demonstrating that the on-shell equation for the string states is
the same as the 't Hooft equation. Since we also get the same integral
equation, such a string interpretation also holds in our case. This is
surprising since we do not have an obvious genus expansion for Feynman
diagrams like in the large $N$ limit.

In Section \ref{majorana}, we treat Majorana quarks in integer spin
$J$ representations of $\text{SU}(2)$. To obtain the meson spectrum in
this case, we apply the Majorana condition on the bilocal fields
discussed above. This projects out half the solutions of the integral
equation and gives a spectrum that is asymptotically linearly spaced,
but with the spacing being twice of that of the Dirac fermion case. In
the case where the quark bare mass is zero, the same condition
projects out the zero mass bound state in the spectrum of the 't Hooft
integral equation. We confirm that there is no meson bound state with
zero mass by solving the equation numerically.

In Section \ref{KLspec}, we work in an axial gauge $A_1 = 0$ and
formulate the theory in terms of collective fields, one of which is
the charge density $\rho_j(x) = \psi^\dag_j(x) \psi^j(x)$, or its
large $J$ version
$\rho(\theta,x) = \psi^\dag(\theta,x)\psi(\theta,x)$. The other
collective field $\sigma(\theta,x)$ appears as a fictitious
$\text{U}(1)$ gauge field minimally coupled to the fermions. The
effective action is rewritten in terms of a scalar $\phi$ with
$\rho = \partial_1\phi$, and the curvature $\eta = \partial_1\sigma$
of the fictitious gauge field $\sigma$. In the large $J$ limit, we
focus on a translation-invariant saddle point of the effective action
at which $\phi$ and $\eta$ are constant. The fermions appear
quadratically and can be integrated out and the resulting
log-determinant is the Euler-Heisenberg effective action for the
fictitious field strength $\eta$. We are not able to solve the saddle
point equations. However, assuming that there is a reasonable solution
for $\eta$ and $\phi$ at the large $J$ saddle point, we are able to
compute the spectral function for the gauge invariant fermion
two-point function and demonstrate that there is no pole in the
two-point function.

The paper also contains several appendices which supplement the main
text with details of the various calculations.

\section{$\text{SU}(2)$ gauge theory with spin $J$ matter}\label{mainsec}

\subsection{The theory}\label{Dirachalfint}

As described in the introduction, we consider $\text{SU}(2)$ gauge
theory in $1+1$ dimensions with a matter quark in the spin $J$
representation of $\text{SU}(2)$.\footnote{The terminology `spin' for
  the gauge representation of $\text{SU}(2)$ should not be confused
  with Lorentz spin since there is no notion of a Lorentz spin in
  $1+1$ dimensions.} It is a well-known fact that irreducible
representations of $\text{SU}(2)$ are real / pseudoreal when $J$ is
integer / half-integer respectively. Thus, when the quark is in an
integer spin representation of $\text{SU}(2)$, the minimal field
content is that of a Majorana spinor. A Dirac spinor in an integer
spin representation is equivalent to two flavours of Majorana spinors
(see Appendix \ref{conventions} for our spinor conventions). For
half-integer spin representations, the single quark must be a Dirac
spinor.\footnote{However, when we have an even number of half-integer
  spin quarks, the pseudoreality can be combined with a
  pseudo-Majorana condition to impose reality properties. For more
  details of the well-known reality properties of fermions charged
  under an $\text{SU}(2)$, we refer the reader to Appendix
  \ref{SU2rep}.} In this section, we work with Dirac fermions in
arbitrary spin representations and treat the case of integer spin
Majorana fermions in Section \ref{majorana}.

The action of our $\text{SU}(2)$ gauge theory with a single Dirac
fermion $\psi$ in a half-integer spin $J$ representation of
$\text{SU}(2)$ is
\begin{equation}\label{mainact}
  \mc{S}[A,\psi] = \int \ud^2x\,\left(-\tfrac{1}{2} \str( F_{\mu\nu} F^{\mu\nu}) + \bar{\psi} \i\slashed{D} \psi +  m \bar\psi\psi\right)\ ,
\end{equation}
where $\ol\psi = \psi^\dag \gamma^0$. The covariant derivative is defined as
\begin{equation}\label{covder}
  (D_\mu \psi)^i = \partial_\mu\psi^i + \i g A^a_\mu (T^a)^i{}_j \psi^j\ ,
\end{equation}
where $g$ is the coupling constant and the generators $T^a$ in
\eqref{covder} are hermitian and are in the spin $J$
representation. The field strength $F_{\mu\nu} = F_{\mu\nu}^a T^a$ in
\eqref{mainact} is written with $T^a$ being matrices in the
fundamental representation of $\text{SU}(2)$; in this case, we have
$\str(T^a T^b) = \frac{1}{2}\delta^{ab}$. The action can be written
explicitly in components as
\begin{align}\label{actJ}
  \mc{S}[A,\psi]
  &=  \int \ud^2x\left(-\tfrac{1}{2} F_{+-}^a F^{+-,a} + 2\i \psi_+^\dag D_- \psi_+ + 2\i \psi^\dag_- D_+ \psi_- + m \psi_+^\dag \psi_- + m \psi_-^\dag \psi_+\right)\ ,
\end{align}
where the vector indices on the field strength and the covariant
derivatives are lightcone indices defined via $x^\pm = x^0 \pm x^1$
(for more details on our notation and conventions, see Appendix
\ref{conventions}). We choose the direction $x^+$ as time. The
component $A_+$ is then non-dynamical and the Gauss' law is the field
equation for $A_+$:
\begin{equation}\label{gaussspinJ}
  G_-^a \equiv (D_-E)^a - 2 g \psi_-^\dag T^a \psi_- = 0\ .
\end{equation}
where the electric field $E^a$ is defined as
$E^a := - F^{+-,a} = 4 F^a_{+-}$. We choose the gauge representative
$A_- = 0$ for the orbits generated by the Gauss' law. In this gauge,
the electric field is given by
\begin{equation}
  E^a = 2g\frac{1}{\partial_-} (\psi^\dag_- T^a \psi_-)\ .
\end{equation}
The component $\psi_+$ is also non-dynamical since its action does not
involve a $\partial_+$ derivative. Its equation of motion is a
constraint:
\begin{equation}\label{psipeom}
  2\i D_-\psi_+ + m \psi_- = 0 \quad\longrightarrow\quad \psi_+ = \frac{\i m}{2} \frac{1}{\partial_-}\psi_-\ \ \text{in}\ \ \text{$A_- = 0$ gauge.}
\end{equation}
When $m = 0$, the $\psi_+$ fermions are no longer related to $\psi_-$,
and decouple in this gauge. We take the Hamiltonian to be the Noether
charge for time translations:
\begin{align}
  H &= P_+ = \frac{1}{2}\int \ud x^- \left(\frac{1}{8} E^a E^a - \frac{m}{2}(\psi_-^\dag\psi_+ + \psi_+^\dag \psi_-)\right)\ ,\nonumber\\
    &= \int \ud x^- \left(-\frac{g^2}{4}  (\psi^\dag_-T^a \psi_-)\frac{1}{\partial_-^2} (\psi^\dag_-T^a \psi_-) - \frac{\i m^2}{4} \psi_-^\dag \frac{1}{\partial_-}\psi_-\right)\ .
\end{align}
(in deriving the above form, we have used the Gauss' law
\eqref{gaussspinJ} and the $\psi_+$ field equation \eqref{psipeom} in
the $A^a_- = 0$ gauge.) We use the appropriate
Fourier transforms in Appendix \ref{conventions} to rewrite the
Hamiltonian as
\begin{multline}\label{hamil}
  H = \int \ud x^- \ud y^- \Big(-\frac{g^2}{8} (\psi^\dag_- T^a \psi_-)(x) |x^- - y^-| (\psi^\dag_- T^a \psi_-)(y) \\ -\frac{\i m^2}{8} \psi^\dag_-(x)\sgn(x^- - y^-) \psi_-(y)\Big)\ .
\end{multline}

\paragraph{The colour-singlet condition:} The Gauss' law in $A_- = 0$
gauge can be integrated over $x^-$ to get
\begin{equation}\label{intgauss}
  E^a(\infty) - E^a(-\infty) =  2g \int \ud x^- \psi_-^\dag T^a \psi_-(x)\ .
\end{equation}
A state $|\Psi\rangle$ is a colour-singlet when the
$E^a(\infty) = E^a(-\infty)$ on the state. From the integrated Gauss'
law \eqref{intgauss}, this is the same as requiring
\begin{equation}\label{colorsing}
  \int \ud x^- \bm{:}\!\psi^\dag(x) T^a \psi(x)\!\bm{:} |\Psi\rangle = 0\ ,
\end{equation}
where the current $\psi^\dag T^a\psi$ is normal-ordered.

\subsection{The large $J$ limit}\label{largeJsec}

Next, we would like to study the large $J$ limit of the above
model. Let us focus on the current $\psi^\dag_- T^a \psi_-(x)$ that
appears in the four-fermi term in the Hamiltonian
\eqref{hamil}. Expanding the fermion $\psi_-$ in the $|J,m\rangle$
basis as $\psi_- = \sum_m\psi^m_-(x) |J,m\rangle$, the components of
the current are
\begin{align}\label{melem}
  &\psi^\dag_- T^+ \psi_-(x) = \sum_{m=-J}^J\sqrt{(J-m)(J+m+1)}\, \psi^*_{m+1,-}(x) \psi^{m}_-(x)\ ,\nonumber\\
  &\psi^\dag_- T^- \psi_-(x) = \sum_{m=-J}^J\sqrt{(J+m)(J-m+1)}\, \psi^*_{m-1,-}(x) \psi^{m}_-(x)\ ,\nonumber\\
  &\phantom{\psi^\dag_- T^- \psi_-(x)} = \sum_{m=-J}^{J}\sqrt{(J-m)(J+m+1)}\, \psi^*_{m,-}(x) \psi^{m+1}_-(x)\ ,\nonumber\\
  &\psi^\dag_- T^3 \psi_-(x) = \sum_m m\, \psi^*_{m-}(x) \psi^{m}_-(x)\ ,
\end{align}
where we have used the standard expressions for the matrix elements of
the generators $T^\pm$ and $T^3$ in the $|J,m\rangle$ basis of the
spin $J$ representation of $\text{SU}(2)$ (see \eqref{Tcomp} in
Appendix \ref{SU2rep} for a recap).

We begin with the following manipulation. Multiply and divide the
coefficients in \eqref{melem} by the square root of the Casimir
$C_2(J) = J(J+1)$ of the spin $J$ representation, and write the
expressions in terms of the following quantities:
\begin{align}\label{J1clebsch}
  &\sqrt{\frac{(J-m)(J+m+1)}{J(J+1)}}\ ,\quad  \frac{m}{\sqrt{J(J+1)}}\ .
\end{align}
In the large $J$ limit, the expression $\sqrt{J(J+1)}$ is approximated
by $J$ at leading order. Thus, at leading order in large $J$, the
above expressions are approximated by
\begin{align}\label{J1clebsch1}
  & \sqrt{\frac{(J-m)(J+m+1)}{J(J+1)}} \approx\sqrt{1-\frac{m^2}{J^2}}\ ,\quad  \frac{m}{\sqrt{J(J+1)}} \approx \frac{m}{J}\ .
\end{align}
Since the range of $m$ is $m =-J,-J+1,\ldots,J-1,J$, the range of
$u_m := m/J$ is from $-1$ to $1$ in steps of $1/J$. In the large $J$ limit,
this range becomes a continuum from $-1$ to $1$ and it is appropriate
to define angles $\theta_m \in [0,\pi]$ by
\begin{equation}\label{thetadef}
  \cos\theta_m := u_m = \frac{m}{J}\ ,\quad \sin\theta_m = \sqrt{1 - \frac{m^2}{J^2}}\ .
\end{equation}
\footnote{In fact, this is nothing new. In the large $J$ limit, the
  spin $J$ representation can be treated semiclassically as a vector
  in three dimensions with length $J$ (actually $\sqrt{J(J+1)}$ which
  is approximately $J$ in the large $J$ limit). The $z$-component of
  this vector is simply $J \cos\theta$ where $\theta$ is the azimuthal
  angle on the sphere
  \cite{wigner2012group,Edmonds+1957,Brussaard1957}. The quantum
  mechanical nature of the spin leads to a discretization of the
  possible $z$-component values at $\theta = \theta_m$ which becomes
  closer and closer to a continuum as we take $J \to \infty$. This can
  also be observed in the coherent state representation of
  $\text{SU}(2)$ in which a coherent state $|\theta,\varphi\rangle$ is
  labelled by the two angles $\theta \in [0,\pi]$ and
  $\varphi \in [0,2\pi]$ which coordinatize the coset
  $\text{SU}(2) / \text{U}(1) = S^2$ \cite{perelomov1977}. The
  expectation value of $T^3$ in the coherent state
  $|\theta,\varphi\rangle$ is simply
  $\langle \theta,\varphi | T^3 |\theta,\varphi \rangle = J
  \cos\theta$. Since coherent states are the closest one can come to
  classical configurations, the above expectation value also justifies
  the definition \eqref{thetadef} in terms of a `classical' angle
  $\theta \in [0,\pi]$ when $J$ is taken to be very large.} The
bilinears $\psi^\dag(x)T^a\psi(x)$ become, at leading order in large
$J$,
\begin{align}\label{melemlargeJ}
  &\psi^\dag_- T^+ \psi_-(x) \to \sqrt{C_2(J)} \sum_m \sin\theta_m \psi^*_{m+1,-}(x) \psi^{m}_-(x)\ ,\nonumber\\
  &\psi^\dag_- T^- \psi_-(x) \to \sqrt{C_2(J)} \sum_m \sin\theta_m \psi^*_{m,-}(x) \psi^{m+1}_-(x)\ ,\nonumber\\
  &\psi^\dag_- T^3 \psi_-(x) \to \sqrt{C_2(J)} \sum_m \cos\theta_m \psi^*_{m-}(x) \psi^{m}_-(x)\ ,
\end{align}
where $\sqrt{C_2(J)} = \sqrt{J(J+1)} \sim J$ in the large
$J$ limit.

\subsection{The continuum field $\psi_-(\theta,x)$}\label{contfield}

The values $u_m = m/J$ lie in the interval $[-1,1]$ and become closer
and closer as $J$ becomes large. Thus, they serve to discretize the
interval $[-1,1]$ in steps of $1/J$. Consequently, the fermion
components $\psi^m_-(x)$ also coalesce into a function
$\tl\psi_-(u,x)$ on the $u$-interval $[-1,1]$. For purposes which will
be clear later, we define the function $\tl\psi_-(u,x)$ with an
additional prefactor $\sqrt{J}$:
\begin{equation}\label{psitldef}
  \tl\psi_-\left(u = \frac{m}{J},x\right) = \sqrt{J}\,\psi^m_-(x)\ .
\end{equation}
As $J$ goes to $\infty$, the fields $\tl\psi_-(\frac{m}{J},x)$ indeed
cluster together on the $u$ interval $u \in [-1,1]$. This enables us
to study the variation of $\tl\psi_-(u,x)$ w.r.t.~small changes in $u$
which are changes of order $1/J$ in the large $J$ limit.

The prefactor of $\sqrt{J}$ in \eqref{psitldef} can be motivated as
follows. The fermion components $\psi^m_-$ are discrete for finite $J$
and the neighbouring components need not be close in any sense since
we integrate over all possible configurations of $\psi^m_-$ in the
path integral. In the large $J$ limit, we have defined the fields $\tl\psi_-(\frac{m+1}{J},x)$ and $\tl\psi_-(\frac{m}{J},x)$ such that
\begin{equation}\label{psidiffeq}
 \frac{1}{\sqrt{J}} \left[\tl\psi_-\left(\frac{m+1}{J},x\right) - \tl\psi_-\left(\frac{m}{J},x\right)\right] = \psi^{m+1}_-(x) - \psi^m_-(x)\ .
\end{equation}
For large $J$, two things happen to the left-hand side:
\begin{enumerate}
\item The two fields are situated infinitesimally close to each other in the $u$ direction, and
\item The difference in the two fields is suppressed by the large $J$ factor.
\end{enumerate}
Thus, in the large $J$ limit, we assume that the difference
$\psi^{m+1}_-$ and $\psi^m_-$ is parametrically small, which in turn
implies the existence and continuity of the field
$\tl\psi_-(u,x)$. Later, this assumption will turn out to be justified
since there will be a smooth solution for $\tl\psi(u,x)$ at the large
$J$ saddle point. The continuity in the variable $u$ will also allow
us to conceive derivatives of $\tl\psi_-(u,x)$ in the $u$ direction.

Another justification for the factor of $\sqrt{J}$ in the definition
of $\tl\psi_-(u,x)$ is as follows. Upon canonical quantization, the
fermions $\psi^m_-(x)$ satisfy the equal-time anticommutation
relations
\begin{equation}
  \{\psi^m_-(x), \psi^*_{n-}(y)\} = \delta^m_n \delta(x^- - y^-)\ .
\end{equation}
In terms of $\tl\psi_-(u,x)$ we have
\begin{equation}\label{intercomm}
\frac{1}{J}  \{\tl\psi_-(u_m,x), \tl\psi^*_-(u_n,x)\} = \delta^m_n \delta(x^- - y^-)\ .
\end{equation}
In the large $J$ limit, the Kronecker delta $\delta^m_n$ goes over to
a Dirac delta function in the $u$ variable in the continuum limit
$J \to \infty$ as
\begin{equation}
  \delta^m_n\ \longrightarrow\ \frac{1}{J}\, \delta(u_m-u_n)\ ,
\end{equation}
so that the continuum limit of \eqref{intercomm} becomes
\begin{equation}\label{contcomm}
  \{\tl\psi_-(u,x), \tl\psi^*_-(u',y)\} = \delta(u-u') \delta(x^- - y^-)\ .
\end{equation}
Note that without the prefactor $\sqrt{J}$ in the definition
\eqref{psitldef}, the anticommutation relations \eqref{contcomm} would
have a $1/J$ on the right-hand side; the right hand side would then go
to zero in the large $J$ limit. The explicit prefactor $\sqrt{J}$ in
the definition of the continuum field $\tl\psi_-(u,x)$ ensures that
the quantum nature of the fermions persists in the large $J$ limit.

We finally make a transformation from the variable $u$ to the angle
$\theta$ related as $u = \cos\theta$ and define
\begin{equation}
  \psi_-(\theta,x) := \tl\psi_-(u,x)\Big|_{u = \cos\theta}\ ,
\end{equation}
where we have omitted the $\tl{}$ in the definition of
$\psi_-(\theta,x)$ on the left-hand side to avoid cluttering of
notation. However, we must emphasize that the scaling $m \to m/J$
persists in the definition of $\psi_-(\theta,x)$ which in turn allows
us to study derivatives w.r.t.~$\theta$ as alluded to earlier in this
subsection.

The equal time anticommutation relations \eqref{contcomm} become
\begin{equation}\label{contcommfin}
  \{\psi_-(\theta,x), \psi^*_-(\theta',y)\} = \frac{1}{\sin\theta}\delta(\theta-\theta') \delta(x^- - y^-)\ .
\end{equation}
Thus, in the continuum limit, the fermion $\psi^m_-(x)$ will be
replaced by
\begin{equation}\label{thetarepl}
  \psi^m_-(x)\quad \longrightarrow \quad \frac{1}{\sqrt{J}} \psi_-(\theta,x)\ .
\end{equation}
The sum over $m$ is also similarly replaced by an integral over $u$, or an integral over $\theta$:
\begin{equation}\label{riemannsum}
  \sum_{m=-J}^J = J\sum_{m=-J}^J \frac{1}{J} \to J \int_{-1}^1 \ud u = J \int_0^\pi \ud\theta \sin\theta\ .
\end{equation}

\subsection{A derivative along the $\theta$ direction}\label{thetader}

Let us look at the fermion bilinear $\psi_-^\dag T^- \psi_-$ that
appears in the four-fermi term in the Hamiltonian:
\begin{equation}
  \psi_-^\dag T^- \psi_- = \sum_{m=-J}^J \sqrt{(J-m) (J+m+1)} \psi^*_{m,-} \psi^{m+1}_-\ .
\end{equation}
Since the term relates the neighbouring components $\psi^m$ and
$\psi^{m+1}$, we expect such a coupling to involve a derivative in the
continuum variable $u = m/J$, or equivalently $\theta = \cos^{-1}
u$. We write $\psi^{m+1}_- = (\psi^{m+1}_- - \psi^{m}_-) + \psi^m_-$
and focus on the difference $\psi^{m+1}_- - \psi^{m}_-$. In terms of
$\tl\psi_-(u,x)$, we have
\begin{align}\label{findiff1}
  \psi^{m+1}_-(x) -   \psi^{m}_-(x)
  &= \frac{1}{\sqrt{J}}\left[\tl\psi_-\left(\frac{m+1}{J},x\right) - \tl\psi_-\left(\frac{m}{J},x\right)\right]\nonumber\\
  &=  \frac{1}{\sqrt{J}}\times\frac{1}{J} \times \frac{\displaystyle\tl\psi_-\left(\frac{m+1}{J},x\right) - \tl\psi_-\left(\frac{m}{J},x\right)}{\displaystyle\frac{m+1}{J} - \frac{m}{J}}\ .
\end{align}
In the large $J$ limit, the above simply becomes the derivative of
$\tl\psi_-(u,x)$ along the $u$ direction:
\begin{equation}
  \psi^{m+1}_-(x) -   \psi^{m}_-(x)\ \ \longrightarrow\ \ \frac{1}{\sqrt{J}} \frac{1}{J} \frac{\partial}{\partial u} \tl\psi_-(u,x)\ .
\end{equation}
That the derivative exists rests on the assumption that the difference
between $\psi^{m+1}_-$ and $\psi^m_-$ is suppressed as $1 / \sqrt{J}$;
see the comments after \eqref{psidiffeq}.

In terms of the $\theta$ coordinate, we have \eqref{thetarepl},
\begin{equation}
  \psi^{m+1}_-(x)  -   \psi^{m}_-(x)\ \ \longrightarrow\ \ -\frac{1}{\sqrt{J}}\frac{1}{J \sin\theta} \frac{\partial}{\partial \theta} \psi_-(\theta,x)\ .
\end{equation}
Thus, in the large $J$ limit,
\begin{equation}\label{thetadiffrepl}
  \psi^{m+1}_-(x)\ \ \longrightarrow\ \ \frac{1}{\sqrt{J}} \left[ \psi(\theta,x) - \frac{1}{J\sin\theta} \frac{\partial}{\partial \theta} \psi(\theta,x)\right]\ ,
\end{equation}
so that
\begin{equation}
  \psi^*_{m-}(x) \psi^{m+1}_-(x)\ \ \longrightarrow\ \ \frac{1}{J}\left[ \psi^*(\theta,x) \psi(\theta,x) - \frac{1}{J\sin\theta} \psi^*(\theta,x)\frac{\partial}{\partial \theta} \psi(\theta,x)\right]\ .
\end{equation}

\subsection{The Hamiltonian in the large $J$
  limit}\label{largeJsummary}
Recall the continuum expressions for the coefficients \eqref{thetadef},
\begin{equation}\label{clebschJcont}
  \sqrt{(J-m)(J+m+1)}  \to  J\sin\theta_m\ ,\quad m \to J \cos\theta_m\ ,
\end{equation}
and those for the fermion fields \eqref{thetarepl} and \eqref{thetadiffrepl},
\begin{equation}\label{psiJcont}
  \psi^m_-(x) \to \frac{1}{\sqrt{J}}\psi_-(\theta,x)\ ,\quad \psi^{m+1}_-(x)  - \psi^m_-(x) \to -\frac{1}{\sqrt{J}} \frac{1}{J\sin\theta}\partial_\theta \psi_-(\theta,x)\ ,
\end{equation}
and for the sum over the representation index \eqref{riemannsum}
\begin{equation}\label{riemannsumJ}
  \sum_m \to J \int_0^\pi \ud\theta\sin\theta\ .
\end{equation}
Using the formulas \eqref{clebschJcont}, \eqref{psiJcont} and
\eqref{riemannsumJ} above, the fermion current can be written in the
continuum $J\to\infty$ limit as
\begin{align}\label{melemJcont}
  &\psi^\dag_- T^+ \psi_-(x) \to  J \int_0^\pi\ud\theta\sin\theta \Big(-\tfrac{1}{J} \partial_\theta\psi^*_-(\theta,x)\, \psi_-(\theta,x) + \sin\theta\psi^*_-(\theta,x) \psi_-(\theta,x)\Big) \ ,\nonumber\\
  &\psi^\dag_- T^- \psi_-(x) \to J \int_0^\pi\ud\theta\sin\theta \Big(-\tfrac{1}{J} \psi^*_-(\theta,x) \partial_\theta\psi_-(\theta,x) + \sin\theta\psi^*_-(\theta,x) \psi_-(\theta,x)\Big)\ ,\nonumber\\
  &\psi^\dag_- T^3 \psi_-(x) \to J \int_0^\pi\ud\theta\sin\theta \cos\theta\, \psi^*_{-}(\theta,x) \psi_-(\theta,x)\ .
\end{align}
Observe that we can drop the $\partial_\theta$ terms in the large $J$
limit since they are suppressed by $1/J$. In principle, we can compute
subleading terms in an $1/J$ expansion by retaining these terms.

At leading order in $1/J$, Hamiltonian \eqref{hamil} can be written in
a simple form where there is a potential in the $\theta$ direction:
\begin{align}\label{hamil2}
  H&=\int \ud x^- \ud y^- \bigg[ -\frac{\i m^2}{8} \int_0^\pi\ud\theta\sin\theta\, \sgn(x^- - y^-)\psi^*_-(\theta,x) \psi_-(\theta,y)\nonumber\\
  &\quad-\frac{g^2}{8}J^2 \int_0^\pi\ud\theta\sin\theta\int_0^\pi\ud\theta'\sin\theta'  |x^- - y^-| \cos(\theta-\theta') \big(\psi^*_{-} \psi_-\big)(\theta,x)\,  \big(\psi^*_{-}\psi_-\big)(\theta',y)\bigg]\ .
\end{align}
Similarly, the colour-singlet conditions \eqref{colorsing} on states
$|\Psi\rangle$ become
\begin{align}\label{colorsing1}
 0 &=  \int \ud x^- \int_0^\pi \ud\theta\sin\theta \bm{:}\! \Big(-\tfrac{1}{J}\big(\partial_\theta\psi^*_- \psi_-\big)(\theta,x) + \sin\theta (\psi^*_-\psi_-)(\theta,x)\Big)\!\bm{:}|\Psi\rangle\ ,\nonumber\\
 0 &=  \int \ud x^- \int_0^\pi \ud\theta\sin\theta \bm{:}\!\Big(-\tfrac{1}{J}\big(\psi^*_- \partial_\theta\psi_-\big)(\theta,x) + \sin\theta (\psi^*_-\psi_-)(\theta,x)\Big)\!\bm{:}|\Psi\rangle\ ,\nonumber\\
 0 &= \int \ud x^- \int_0^\pi \ud\theta\sin\theta \cos\theta\, \bm{:}\!(\psi^*_-\psi_-)(\theta,x)\!\bm{:}|\Psi\rangle\ .
\end{align}

\section{Mesons and $W_\infty$ coadjoint orbits}\label{thooftsec}

\subsection{A 't Hooft-like limit}
We reproduce the Hamiltonian of our theory \eqref{hamil}:
\begin{multline}\label{hamil1}
  H  = \int \ud x^- \ud y^- \bigg(-\frac{g^2}{8} (\psi^\dag_- T^a \psi_-)(x) |x^- - y^-| (\psi^\dag_- T^a \psi_-)(y) \\ -\frac{\i m^2}{8} \psi^\dag_-(x)\sgn(x^- - y^-) \psi_-(y)\bigg)\ .
\end{multline}
Let us focus on the mass term
\begin{multline}
  -\frac{\i m^2}{8} \psi^\dag_-(x) \sgn(x^- - y^-) \psi_-(y) =  \frac{\i m^2}{8}  \sgn(x^- - y^-) \sum_{n=-J}^J  \psi^n_-(y) \psi^*_{n-}(x) \\ =\frac{\i m^2}{8}  \sgn(x^- - y^-)\, \str \Big( \psi_-(x)  \psi^\dag_-(y)\Big) \ ,
\end{multline}
where the trace $\str$ is simply the sum over the representation
indices as indicated. As is familiar from studies of the large $N$
limit of quantum field theories, the trace is expected to scale as
$2J+1$ in the large $J$ limit since it has $2J+1$ terms. Thus, the
first observation in the large $J$ limit is that
\begin{enumerate}
\item Every trace over representation indices is $\mc{O}(J)$ in the
  large $J$ limit.
\end{enumerate}
It is easy to see that the Hamiltonian can be written as a single
trace over the representation indices. To see this, and for subsequent
calculations, we introduce the bilocal field
\begin{equation}
  M(x,y;\theta,\theta') := \psi_-(\theta,x) \psi^*_-(\theta',y)\ ,
\end{equation}
or, in more detail,
\begin{equation}\label{Mdef}
  M(x^+,x^-,y^+,y^-;\theta,\theta') := \psi_-(\theta,x^+,x^-) \psi^*_-(\theta',y^+,y^-)\ .
\end{equation}
The Hamiltonian \eqref{hamil2} can now be written as a single trace
over the gauge representation indices (which are parametrized by the
angle $\theta$ in the continuum limit):
\begin{equation}\label{hamilfin}
  H = \int \ud x^- \ud y^- \bigg[ \frac{\i m^2}{8}  \sgn(x^- - y^-)\, \str\, M(y,x)  + \frac{g^2 J^2}{8}  |x^- - y^-|\, \str\big( M(y,x) \wt{M}(x,y)\big)\bigg]\ .
\end{equation}
where
$\wt{M}(x,y;\theta,\theta') = \cos(\theta-\theta')
M(x,y;\theta,\theta')$, and the bilocal fields are all at equal time
$x^+ = y^+$. Since the Hamiltonian is a single trace over the
representation indices, we have our second observation:
\begin{enumerate}
\item[2.] The Hamiltonian is $\mc{O}(J)$ in the large $J$ limit
  provided the combination $g^2 J^2$ is kept fixed in the limit.
\end{enumerate}
Again, in analogy with the large $N$ limit of gauge theories, We
define the 't Hooft-like coupling
\begin{equation}\label{thooftcoup}
  \lambda = g^2 J^2\ ,
\end{equation}
and work in the limit
\begin{equation}
\boxed{g^2 \to 0\ ,\quad J \to \infty\ ,\quad \text{with}\ \ \lambda = g^2 J^2\ \ \text{fixed.}}
\end{equation}
Since the Hamiltonian is $\mc{O}(J)$ in the 't Hooft-like large $J$
limit, and it is expressed completely in terms of the bosonic bilocal
variables $M$, we can treat $M$ semiclassically in this limit. The
advantage of this formulation is that one may be able to directly
extract the meson, i.e., gauge invariant quark-antiquark, spectrum and
wavefunctions from a saddle-point analysis of the bilocal variables.

The above idea was successfully applied to the large $N$ limit of
$\text{SU}(N)$ QCD in $1+1$ dimensions coupled to quarks in the
fundamental representation \cite{Dhar:1994ib}, where it was used to
derive the integral equation for the quark-antiquark wavefunction
first obtained by 't Hooft \cite{tHooft:1974pnl}. In this paper, we
follow the steps of \cite{Dhar:1994ib} closely in spirit but keeping
one crucial difference in mind: we work with bilocal operators
$M(x^-,y^-;\theta,\theta')$ that are non gauge-singlets whereas the
bilocal operators in \cite{Dhar:1994ib} were gauge-invariant. At the
end, we perform the additional step of extracting the gauge invariant
information from our results.

We also note an identity satisfied by the equal-time bilocal operators
which will be useful for our semiclassical analysis of the model in
the 't Hooft-like large $J$ limit:
\begin{equation}\label{idempot}
  \int \ud z^- \int_0^\pi\ud\varphi\sin\varphi\, M(x^-,z^-;\theta,\varphi)M(z^-,y^-;\varphi,\theta') = M(x^-,y^-;\theta,\theta') \big(1 + Q\big)\ ,
\end{equation}
\footnote{This can be seen by the following manipulation. The left hand side is
\begin{align}
%  &\int \ud z^- \int_0^\pi\ud\varphi\sin\varphi\, M(x,z;\theta,\varphi)M(z,y;\varphi,\theta')\nonumber\\
   &\int \ud z^- \int_0^\pi\ud\varphi\sin\varphi\, \psi(\theta,x)\psi^*(\varphi,z) \psi(\varphi,z) \psi^*(\theta',y)\nonumber\\ 
   &= \int \ud z^- \int_0^\pi\ud\varphi\sin\varphi\, \psi(\theta,x)\psi^*(\varphi,z)\left[ \frac{1}{\sin\varphi}\delta(\varphi-\theta')\delta(z^--y^-) -  \psi^*(\theta',y)\psi(\varphi,z)\right]\ ,\nonumber\\
%  &= \psi(\theta,x)\psi^*(\theta',y) + \psi(\theta,x) \psi^*(\theta',y) \int \ud z^- \int_0^\pi\ud\varphi\sin\varphi\, \psi^*(\varphi,z)\psi(\varphi,z)\ ,\nonumber\\
   & = M(x,y;\theta,\theta') \big(1 + Q\big)\ .\nonumber
\end{align}
}where $Q$ is the conserved global $\text{U}(1)$ charge operator
supported on a constant time slice $x^+ = \text{const.}$:
\begin{align}\label{conscharge}
  Q &= \int \ud x^- \sum_m \psi^*_{m-}(x^+,x^-) \psi^m_-(x^+,x^-)\nonumber \\
    &= \int \ud x^- \int_0^\pi \ud\theta\sin\theta\,\psi^*_-(\theta,x^+,x^-) \psi_-(\theta,x^+,x^-)\ .
\end{align}
The charge $Q$ is nothing but the baryon number which measures the
number of gauge invariant states of the form
$\bm{:}\!\varepsilon_{i_1\cdots i_{2J+1}}\psi^{i_1}\cdots
\psi^{i_{2J+1}}\!\bm{:}(x)$ where $\varepsilon_{i_1\cdots i_{2J+1}}$
is the Levi-Civita symbol in $2J+1$ dimensions. Note that such a
charge operator does not exist when the fermions $\psi^i$ transform in
the Majorana representation for integer $J$. Indeed, using the
Majorana condition $\psi^*_m = (-1)^{J+m} \psi^{-m}$ (see Appendix
\ref{SU2rep}), we see that the charge operator is simply zero since
the terms corresponding to $-m$ and $m$ in the summation cancel each
other and the $m = 0$ term vanishes after normal ordering since
$\bm{:}\!(\psi^0)^2\!\bm{:} = 0$.

\subsection{The $W_\infty$ algebra and coadjoint orbits}\label{Winfalg}

Next, we show that the quantum operators $M(x^-,y^-;\theta,\theta')$
satisfy a Lie algebra which is the $W^{2J+1}_\infty$ algebra. Recall
that the continuum fermions $\psi_-(\theta,x)$ satisfy the equal time
anti-commutation relations \eqref{contcommfin} which we reproduce here
for convenience:
\begin{equation}\label{ACR}
  \{\psi_-(\theta,x^+,x^-),\psi^*_-(\theta',x^+,y^-) \} =\frac{1}{\sin\theta} \, \delta(\theta-\theta') \, \delta(x^- - y^-)\ .
\end{equation}
Using \eqref{ACR}, it is easy to see that the equal-time
bilocal variables satisfy
\begin{multline}\label{bial}
  [M(x_1^{-},y_1^{-};\theta_1,\theta'_1),M(x_2^{-},y_2^{-};\theta_2,\theta'_2)] \\= \frac{1}{\sin\theta_2} \, \delta(\theta'_1-\theta_2) \delta(y_1^{-} - x_2^{-}) M(x_1^{-},y_2^{-};\theta_1,\theta'_2) \\
  - \frac{1}{\sin\theta_1} \, \delta(\theta_1-\theta'_2) \delta(x_1^{-} - y_2^{-}) M(x_2^{-},y_1^{-};\theta_2,\theta'_1)\ .
\end{multline}
The above algebra of the equal-time bilocal variables
$M(x^-,y^-;\theta,\theta')$ is the 
$W_\infty^{2J+1} := W_\infty \times \text{U}(2J+1)$ algebra (see
\cite{Dhar:1994ib,Dhar:1992hr,Dhar:1993jc}). We next obtain a kinetic
term for the bilocal fields $M$ by appealing to the theory of
quantization of coadjoint orbits of the $W_\infty$ algebra. The orbit
method has found many applications in physical problems, such as
understanding the non-relativistic fermion description of the $c=1$
matrix model \cite{Dhar:1993jc,Dhar:1992hr,Das:1991uta,Dhar:1992rs},
the 't Hooft model of two dimensional QCD with fundamental quarks
\cite{Dhar:1994ib}, gravity duals for the SYK model \cite{Mandal:2017thl},
a non-linear bosonization of fermi surfaces \cite{Delacretaz:2022ocm},
and so on.

Following previous treatments in
\cite{Dhar:1994ib,Das:1991uta,Dhar:1992hr,Dhar:1992rs,Dhar:1993jc}, we
describe the coadjoint orbit of the $W^{2J+1}_\infty$ algebra which is
relevant for us. We only give a summary here and refer the reader to
Appendix \ref{coadj} for more details. The coadjoint orbit is described by
certain constraints on the bilocal field
$M(x^-,y^-;\theta,\theta')$. To describe these constraints, it is
convenient to introduce a matrix notation for the bilocal
fields. Since we have one field for each $(x^-,y^-,\theta,\theta')$,
we can write down a matrix ${\mM}$ such that
\begin{equation}\label{Mmatrix}
  {M}(x^-,y^-;\theta,\theta') = \langle x^-,\theta| {\mM}| y^-,\theta'\rangle\ .
\end{equation}
In terms of ${\mM}$, the constraints that describe the coadjoint orbit
are
\begin{equation}\label{olMconst}
  {\mM}{}^2 = {\mM}\ ,\quad \tr(\bm{1} - {\mM}) = {Q}\ ,
\end{equation}
where $\tr$ is the trace over both the representation indices and
spatial labels $x,y,\ldots,$ and ${Q}$ is the baryon number charge
defined in \eqref{conscharge}.\footnote{Regarding the notation,
  ${\mM}{}^2$ corresponds to matrix multiplication in the
  $(x^-,\theta)$ space, the $\bm{1}$ is the identity matrix in
  $(x^-,\theta)$ space and the $\str$ is a trace over the $x^-$ and
  $\theta$ labels.}  The coadjoint orbit we choose corresponds
choosing $Q = 0$. The above constraints can be shown to be invariant
under the action of the $W^{2J+1}_\infty$ algebra on the ${\mM}$.

\subsection{The action for ${M}$ and the large $J$ saddle point
  equations}\label{Mact}

Next, we describe an action for the classical field ${\mM}$ based
on the discussion above. First, the Hamiltonian for ${\mM}$ is
nothing but the Hamiltonian which we derived from the fermion field
theory \eqref{hamilfin} in terms of the bilocal variables:
\begin{equation}\label{hamilcoadj}
  H = \tr\left[\frac{\i m^2}{8} \mS {\mM} + \frac{\lambda}{8} {\mM} \wt{\mM}\right]\ .
\end{equation}
where the trace $\tr$ is in the $(x,\theta)$ space, and we have
defined $\mS$ and $\wt{{\mM}}$ that have the matrix elements
\begin{align}
  \langle x,\theta | \mS | y,\theta'\rangle
  &= \sgn(x^- - y^-) \frac{1}{\sin\theta}\delta(\theta-\theta')\ ,\nonumber\\
  \langle x,\theta | \wt{\mM}| y,\theta' \rangle
  &= |x^- - y^-| \cos(\theta-\theta') {M}(x,y;\theta,\theta')\ .
\end{align}
The `kinetic' term for the ${\mM}$ can be borrowed from the theory
of quantization of coadjoint orbits where it is derived from the
symplectic structure on the coadjoint orbit. It takes the form
\begin{equation}\label{kinetic}
  \i \int_\Sigma \ud s \ud x^+ \tr ({\mM} [\partial_+ {\mM}, \partial_s {\mM}])\ ,
\end{equation}
where $\Sigma$ is the two-dimensional surface with coordinates $s$ and
$x^+$, with $s$ a semi-infinite coordinate, $-\infty < s \leq 0$,
which is customary in the orbit method. The field $\mM$ is a
function of both $s$ and $x^+$ with boundary conditions
\begin{equation}
  {\mM}(s=0,x^+) = {\mM}(x^+)\ ,\quad \lim_{s \to -\infty} {\mM}(s,x^+) = \text{constant in $x^+$}.
\end{equation}
Thus, the total action for the field ${\mM}$ is
\begin{equation}\label{Maction}
  \mc{S}[{\mM}] = \i \int_\Sigma \ud s \ud x^+ \tr ({\mM} [\partial_+ {\mM}, \partial_s {\mM}]) - \int \ud x^+ \tr\left(\frac{\i m^2}{8} \mS {\mM} + \frac{\lambda}{8} {\mM} \wt{{\mM}}\right)\ ,
\end{equation}
where the $\tr$ is on the $x^-$ and $\theta$ labels of
the field ${\mM}$ \eqref{Mmatrix}.

Since the action \eqref{Maction} involves a single trace over the
$\theta$ labels, it is $\mc{O}(J)$ in the large $J$ limit. Thus, the
large $J$ dynamics is described by the saddle points of
$\eqref{Maction}$. To obtain the saddle point equations, we perform an
infinitesimal variation of ${\mM}$ along the coadjoint orbit:
\begin{equation}
  \delta {\mM} = \i [\bm{\epsilon}, {\mM}]\ ,
\end{equation}
that is,
\begin{equation}
  \delta {M}(x,y;\theta,\theta') = \i \int \ud z^- \int_\phi \Big(\epsilon(x,z;\theta,\phi) {M}(z,y;\phi,\theta') - {M}(x,z;\theta,\phi) \epsilon(z,y;\phi,\theta')\Big)\ .
\end{equation}
Under a variation of the above kind, the variation of the action
\eqref{Maction} is
\begin{equation}\label{actvar}
  \delta \mc{S} =  \i  \int_\Sigma \ud s \ud x^+ \tr\left( \partial_s (\i\bm{\epsilon} \partial_+ {\mM}) - \partial_+ (\i\bm{\epsilon} \partial_s {\mM})\right) - \int \ud x^+ \tr\left(\frac{\i m^2}{8} \i\bm{\epsilon} [{\mM}, \mS] + \frac{\lambda}{4}  \i\bm{\epsilon}[ {\mM}, \wt{{\mM}}]\right)\ .
\end{equation}
Ignoring the partial derivative in time $\partial_+$, we get the
equation of motion
\begin{equation}\label{eomcoadj}
  \i \partial_+ {\mM} = \frac{\i m^2}{8} [{\mM}, \mS] + \frac{\lambda}{4} [{\mM}, \wt{{\mM}}]\ .
\end{equation}
Define the Fourier components of ${M}(x^-,y^-;\theta,\theta')$ as
\begin{equation}
  {M}(x^-,y^-;\theta,\theta') = \int \frac{\ud k_-}{2\pi} \frac{\ud k'_-}{2\pi} \e^{-\i k_- x^- + \i k'_- y^-}\, {M}(k_-,k'_-;\theta,\theta')\ .
\end{equation}
Here onward, we drop the $-$ in the superscripts of coordinates and
subscripts of momenta to avoid clutter. All one dimensional integrals
in the equations of motion will refer to positions and momenta in the
spatial directions $x^-$, $k_-$. The equation of motion becomes
\begin{multline}\label{eomcoadjmom}
  \i\partial_+ {M}(p,k) = \frac{m^2}{4} \left(\frac{1}{p} - \frac{1}{k}\right) {M}(p,k) \\ + \frac{\lambda}{2} \int \frac{\ud q}{-2\pi q^2} \int \frac{\ud r}{2\pi} \big( {M}(p,r+q) \wt{{M}}(r, k-q) -  \wt{{M}} (p-q,r-q) {M}(r,k)\big)\ ,
\end{multline}
where
$\wt{{M}}(p,r;\theta,\theta') =
\cos(\theta-\theta'){M}(p,r;\theta,\theta')$, and we have employed
matrix notation for the $\theta,\theta'$ labels but explicitly display
the spacetime momentum labels on ${M}$.

Recall the constraints \eqref{olMconst} for a coadjoint orbit of the
$W^{2J+1}_\infty$ algebra. The coadjoint orbit we choose is the one
corresponding to ${Q} = 0$, i.e., zero $\text{U}(1)$ charge. Then, any
representative ${\mM}$ of the coadjoint orbit has to satisfy
${\mM}^2 = {\mM}$ and $\tr(\bm{1} - \mM) = 0$. The following choice
satisfies this condition:
\begin{equation}
  {M}_0(k_-,k'_-;\theta,\theta') = 2\pi\Theta(k_-)\delta(k_- - k'_-) \frac{1}{\sin\theta}\delta(\theta-\theta')\ .
\end{equation}
See \cite{Dhar:1994ib} for some motivation for this
choice. Fluctuations along the orbit about the classical solution
${\mM}_0$ are parametrized as
\begin{equation}\label{flucM}
  {\mM} = \e^{\i \mW / \sqrt{J}}\, {\mM}_0\, \e^{-\i \mW/\sqrt{J}} = {\mM}_0 + \frac{\i}{\sqrt{J}} [\mW, {\mM}_0] - \frac{1}{2 J} [\mW, [\mW, {\mM}_0]] + \mc{O}(J^{-3/2})\ ,
\end{equation}
where the $1/\sqrt{J}$ is present to indicate that we consider
fluctuations that are subleading in the large $J$ limit. The
fluctuating field $\mW$ is expected to be the quark-antiquark
wavefunction, though one has to take an additional trace over the
$\theta$, $\theta'$ labels to arrive at the gauge invariant
wavefunction. We will perform this step at the very end of this
analysis in Section \ref{thoofteq}.

In terms of spatial momenta, we have
\begin{multline}\label{WMfluc}
  {M}(k,k';\theta,\theta') = 2\pi\Theta(k_-) \delta(k_- - k'_-) \frac{\delta(\theta-\theta')}{\sin\theta} + \frac{\i}{\sqrt{J}} \Big(\Theta(k'_-) - \Theta(k_-)\Big) W(k,k';\theta,\theta') \\ - \frac{1}{2J} \int \frac{\ud p}{2\pi} \int_\varphi \Big(\Theta(k') + \Theta(k) - 2\Theta(p)\Big) W(k,p;\theta,\varphi) W(p,k';\varphi,\theta') + \mc{O}(J^{-3/2}) \ .
\end{multline}
We obtain an action for the fluctuations $\mW$ by substituting the
above ansatz for ${\mM}$ and retaining terms with an explicit $1/J$
in them. Due to the trace $\tr$ in the action \eqref{Maction} which
is of order $J$, the action for $\mW$ will be of order $1$. Following
\cite{Dhar:1994ib}, we split the momentum components of $\mW$ into
four different fields. Let $q$ and $r$ be positive momenta. Then, let
\begin{align}
  &W^{\rm pp}(q,r) = W(q,r)\ ,\quad W^{\rm pn}(q,r) = W(q,-r)\ ,\nonumber\\
  &W^{\rm np}(q,r) = W(-q,r)\ ,\quad W^{\rm nn}(q,r) = W(-q,-r)\ .
\end{align}
The action for $\mW$ is computed in detail in Appendix
\ref{adjorbsim}. The final form of the action for the $W$ modes is
\begin{align}\label{Wflucaction}
  &\frac{1}{J}  \int_{p > 0} \int_{r > 0} \Bigg[-\i\,\str\big(W^{\rm pn} (p,r)\partial_+ W^{\rm np}(r,p)\big) - \frac{m^2}{4}\left(\frac{1}{r} + \frac{1}{p}\right) \str\big(W^{\rm pn}(p,r) W^{\rm np}(r,p)\big)\nonumber\\
  &\qquad\qquad\qquad-\frac{\lambda}{4}  \int_{-p}^r \frac{\ud k}{-2\pi k^2}  \str\bigg[ \left(\wt{W}^{\rm pn}(r-k,p+k) - W^{\rm pn}(r,p)\right)\, W^{\rm np}(p,r) \nonumber\\
  &\qquad\qquad\qquad\qquad\qquad\qquad\qquad+  \left(\wt{W}{}^{\rm np}(p+k,r-k) - W^{\rm np}(p,r)\right) W^{\rm pn}(r,p)\bigg]\Bigg]\ ,
\end{align}
where
$\wt{W}(s,t;\theta,\theta') =
W(s,t;\theta,\theta')\cos(\theta-\theta')$ and $\str$ is the trace
over the $\theta$ labels. Varying the action w.r.t.~$W^{\rm pn}(s,t)$,
we get the following equation of motion for $W^{\rm np}(s,t)$:
\begin{equation}\label{Weom}
  -\i \partial_+ W^{\rm np}(t,s) - \frac{m^2}{4} \left(\frac{1}{s} + \frac{1}{t}\right) W^{\rm np}(t,s) 
  -\frac{\lambda}{4\pi} \int_{-t}^s \frac{\ud k}{-k^2} \Big(\wt{W}{}^{\rm np}(t+k,s-k) - W^{\rm np}(t,s)\Big) = 0\ .
\end{equation}
Displaying the $\theta$ labels explicitly, we get
\begin{multline}\label{Weomexpl}
  -\i \partial_+ W^{\rm np}(t,s;\theta,\theta') - \frac{m^2}{4} \left(\frac{1}{s} + \frac{1}{t}\right) W^{\rm np}(t,s;\theta,\theta') \\
  -\frac{\lambda}{4\pi} \int_{-t}^s \frac{\ud k}{-k^2} \Big(W^{\rm np}(t+k,s-k;\theta,\theta')\cos(\theta-\theta') - W^{\rm np}(t,s;\theta,\theta')\Big) = 0\ .
\end{multline}

\subsection{The 't Hooft equation for the meson
  wavefunction}\label{thoofteq}

Displaying the $x^+$ dependence explicitly for the $W^{\rm np}(t,s)$,
we define the Fourier transform
\begin{equation}
  W^{\rm np}(t,s;x^+) = \int\frac{\ud r_+}{2\pi} \phi(t,s;r_+) \e^{\i r_+ x^+}\ .
\end{equation}
Also, let us define the centre-of-mass momentum $r_-$ and fractional
momenta $x$
\begin{equation}\label{xdef}
r_- = s_- + t_-\ ,\quad x = \frac{t_-}{r_-}\ .
\end{equation}
We write the Fourier mode $\phi(t,s;r_+) = \phi(r_- x, r_-(1-x); r_+)$
as $\phi(x)$ for short:
\begin{equation}
  \phi(t,s;r_+) = \phi(r_- x, r_-(1-x); r_+) \equiv \phi(x)\ .
\end{equation}
Let us also define a new variable of integration:
\begin{equation}
  y = \frac{k_- + t_-}{r_-}\ .
\end{equation}
The equation of motion \eqref{Weom} can be written in terms of these
quantities as
\begin{multline}\label{phieqfin}
  4r_+r_-\phi(x;\theta,\theta')  =  m^2 \left(\frac{1}{1-x} + \frac{1}{x}\right) \phi(x;\theta,\theta') \\
  -\frac{\lambda}{\pi} \int_{0}^1 \frac{\ud y}{(y-x)^2} \Big(\phi(y;\theta,\theta')\cos(\theta-\theta') - \phi(x;\theta,\theta')\Big)\ .
\end{multline}
Define $\Phi(x)$ to be the trace of
$\phi(x,\theta,\theta')$ over the $\theta$ labels:
\begin{equation}\label{ginvtrace}
  \Phi(x) = \int_0^\pi \ud\theta\sin\theta\,\phi(x;\theta,\theta)\ .
\end{equation}
The above trace is gauge invariant since it originates from the gauge-invariant Wilson line in
$A_- = 0$ gauge $\sum_m \psi^m(x) \psi^*_m(y)$. Taking the trace over $\theta$,$\theta'$ in
\eqref{phieqfin} we get
\begin{equation}\label{ginvphieq}
  4r_+r_-\Phi(x)  =  m^2 \left(\frac{1}{1-x} + \frac{1}{x}\right) \Phi(x) 
  -\frac{\lambda}{\pi} \int_{0}^1 \frac{\ud y}{(y-x)^2} \Big(\Phi(y) - \Phi(x)\Big)\ .
\end{equation}
This is the integral equation for the meson wavefunction obtained by 't
Hooft for $\text{SU}(N)$ gauge theory coupled to a fundamental quark
in the large $N$ limit. The wavefunctions satisfy the boundary
conditions
\begin{equation}\label{bcmeson}
  \Phi(x) \sim \left\{\begin{array}{cc} x^\beta & \quad x \to 0\ , \\ (1-x)^\beta &\quad x \to 1\ , \end{array}\right.
\end{equation}
where the exponent $\beta$ satisfies $\beta < 1$ and
\begin{equation}\label{bcthooft}
  \pi\beta\cot\pi\beta + \frac{\pi m^2}{\lambda} = 1\ .
\end{equation}
The above conditions can be obtained by demanding that the right hand
side of \eqref{ginvphieq} is regular as $x \to 0$ and $x \to 1$
respectively (see Appendix \ref{boundary}).

\section{The meson spectrum for integer spin $J$}\label{majorana}

In this section, we analyze the model for integer spin $J$ where the
minimal representation of the quarks is the $2J+1$ dimensional real
(or Majorana) representation of $\text{SU}(2)$. We have described the
representation theory of $\text{SU}(2)$ in Appendix \ref{SU2rep} and
just state the relevant facts here. Irreducible representations of
$\text{SU}(2)$ with integer spin $J$ are real, and hence fermions in
such representations can be taken to be Majorana. The Majorana
condition on a Dirac fermion $\Psi$ in an integer spin $J$
representation is
\begin{equation}\label{realityJm1}
  \Psi^*_{-m,+} = -(-1)^{J+m} \Psi^m_+\ ,\quad  \Psi^*_{-m,-} = (-1)^{J+m} \Psi^m_-\ ,
\end{equation}
where $\Psi^m$ are the components in the $|J,m\rangle$ basis. To
obtain the results for Majorana fermions, we consider the results for
a Dirac fermion in the integer spin $J$ representation and study the
consequences of the reality condition \eqref{realityJm1}.

\subsection{Reality condition on the meson wavefunction}

Recall from the Dirac fermion analysis in Section \ref{thoofteq} that
the gauge-variant wavefunction $\phi(x,\theta,\theta')$ is simply the
$1/J$ fluctuation of the bilocal field written in momentum space
$\Psi_-(\theta,p_-)\Psi^*_-(\theta',q_-)$, with
$x = p_- / (p_- + q_-)$. Let us apply the reality conditions
\eqref{realityJm1} on its discrete version
$\Psi^m_-(p_-) \Psi^*_n(q_-)$:
\begin{multline}
  \Psi^m_-(p_-)\Psi^*_{n-}(q_-) = (-1)^{J+n} \Psi^{m}_{-}(p_-) \Psi^{-n}_-(q_-)\\ = -  (-1)^{J+n} \Psi^{-n}_-(q_-) \Psi^{m}_{-}(p_-)  =  -(-1)^{m+n} \Psi^{-n}_{-}(q_-) \Psi^*_{-m,-}(p_-)\ .
\end{multline}
Since we are interested in the gauge-invariant version
$\Phi(x) = \int\ud\theta\sin\theta\, \phi(x,\theta,\theta)$, we only
need the reality properties on the above bilocals with $m=n$:
\begin{multline}\label{contreal}
  \Psi^m_-(p_-)\Psi^*_{m-}(q_-) = - \Psi^{-m}_{-}(q_-) \Psi^*_{-m,-}(p_-)\\ \Rightarrow\quad \Psi_-(\theta,p_-) \Psi^*_-(\theta,q_-) = - \Psi_-(\pi-\theta,q_-)\Psi^*_-(\pi-\theta,p_-)\ .
\end{multline}
Recall from \eqref{xdef} that $x$ is the fraction of the total
momentum $p_- +q_-$ carried by the first fermion in the bilocal. Thus,
when the two momenta are interchanged $p_- \leftrightarrow q_-$ in
\eqref{contreal}, we have $x \leftrightarrow 1-x$. Thus, the reality
condition \eqref{contreal} on the bilocal translates to the following
on $\phi(x;\theta,\theta)$:
\begin{equation}
  \phi(x;\theta,\theta) = -\phi(1-x;\pi-\theta,\pi-\theta)\ .
\end{equation}
Integrating the above equation over $\theta$, we get the following
condition on the meson wavefunction $\Phi(x)$:
\begin{equation}\label{reflec}
  \Phi(x) = -\Phi(1-x)\ .
\end{equation}
Recall the 't Hooft equation satisfied by $\Phi(x)$:
\begin{equation}\label{majthooft}
  4r_+r_-\Phi(x)  =  m^2 \left(\frac{1}{1-x} + \frac{1}{x}\right) \Phi(x) 
  -\frac{\lambda}{\pi} \int_{0}^1 \frac{\ud y}{(y-x)^2} \Big(\Phi(y) - \Phi(x)\Big)\ .
\end{equation}
The reality condition \eqref{reflec} on $\Phi(x)$ thus dictates that
we must retain only those solutions of \eqref{majthooft} which satisfy
\eqref{reflec}. Let us study the consequences of projecting out the
other solutions. The 't Hooft equation \eqref{majthooft} for the Dirac
fermion has the asymptotic solutions indexed by a positive integer $k$
\cite{tHooft:1974pnl}
\begin{equation}
  \Phi_k(x) \sim \sin \pi k x\ ,\quad \text{with eigenvalue}\quad  M^2_k = \pi\lambda k\ .
\end{equation}
Of the above, only solutions with $k$ even satisfy the reality
condition \eqref{reflec} and hence only those solutions must be
retained. Writing $k = 2\ell$, the asymptotic meson spectrum for a
Majorana fermion is
\begin{equation}
  M_\ell^2 = 2\pi\lambda\ell\ ,\quad \text{with $\ell$ a positive integer.}
\end{equation}
The asymptotic meson spectrum for a Majorana fermion has double the
spacing compared to that of a Dirac fermion.

\subsection{Zero quark mass}

For the case $m = 0$, 't Hooft \cite{tHooft:1974pnl} has demonstrated
that there is a solution of the integral equation \eqref{majthooft}
which has zero eigenvalue $M^2 = 0$ with eigenfunction
$\Phi_0(x) = 1$. Clearly, this is incompatible with the reality
condition \eqref{reflec} and is projected out. However, there could be
zero mass bound states of the integral equation if we choose boundary
conditions compatible with \eqref{reflec}. Indeed, from the boundary
analysis for the $m=0$ case (see eqs.~\eqref{bcmeson},
\eqref{bcthooft} and Appendix \ref{boundary}), it is clear that
$\Phi(x)$ has to approach constants $c_0$, $c_1$ as $x \to 0$ or as
$x \to 1$ respectively. To find if such a zero mass bound state
exists, we appeal to numerical
analysis.

Since the boundary conditions \eqref{bcmeson} tells us that for $m=0$,
the eigenfunction need not vanish at the boundaries, we choose the
ansatz
\begin{equation}
  \Phi(x) = \sum_{n=1}^{K} b_n\cos\big((2n-1)\pi x\big)\ .
\end{equation}
This forms a complete basis for functions satisfying the reality
condition \eqref{reflec} on the interval $[0,1]$. Plugging this ansatz
in the 't Hooft equation \eqref{majthooft} and exploiting the
orthogonality of cosines, we get a matrix equation for the
coefficients $b_n$
\begin{equation}
  M^2 b_k = \sum_n B_{kn} b_n\ ,
\end{equation}
with
\begin{multline}
  B_{kn} = -2\frac{\lambda}{\pi} \int_0^1 \ud x \, \cos\big((2k-1)\pi x\big) \bigg[ \left(\frac{1}{x}+\frac{1}{1-x}\right) \cos\big((2n-1)\pi x\big) \\ + \int_{0}^{1} \frac{\ud y}{(y-x)^2} \cos\big((2n-1)\pi y\big) \bigg]\ .
\end{multline}
The eigenvalues and eigenvectors of the matrix $B$ are then calculated
numerically for increasingly large values of integer $K$ to ensure
convergence. We find that the lowest eigenvalue is non-zero and is
given by $M^2_0 \approx 1.8722\, \lambda$.\footnote{Already
  for $K=10$, the numerical value is stable up to 5 significant
  digits. Increasing to $K=50$, $M^2_0 \approx 1.8722003$ is stable up
  to 7 significant digits.} Therefore, even for $m=0$, the spectrum
of the integral equation \eqref{majthooft} satisfying the condition
\eqref{reflec} is gapped.

\section{Local collective variables}\label{KLspec}

In this section, we analyze the dynamics of the model in terms of
bosonic collective variables
$\rho(\theta,x) = \psi^*(\theta,x) \psi^m(\theta,x)$ and
$\sigma(\theta,x)$ which are local in the coordinates, as opposed to
the bilocal variables $M(x,y)_{ij}$ of the previous sections. The
collective field $\sigma$ acts as an effective electric field for the
fermion -- we assume the electric field to be constant at the large
$J$ saddle point. The fermion determinant is then the standard
Euler-Heisenberg lagrangian for a fermion in a constant electric
field. We then analyze the large $J$ saddle point equations and
compute the spectral function for the gauge-invariant fermion
two-point function and observe that there is no pole in the
propagator. The lack of a pole in the gauge invariant two-point
function signifies that a single fermion is not observed in asymptotic
states, thus reinforcing the picture of quark confinement that emerged
in the discrete meson spectrum computed in the previous sections.

\subsection{The effective action in terms of collective variables}
In this section, we will work with the usual notion of time as
$x^0$. The action can be written explicitly in terms of the gauge
field as
\begin{align}\label{0actJ}
  &\mc{S}[A,\psi]\nonumber\\
  &= \int \ud^2x\, \left(\frac{1}{2} (\partial_0 A^a_1)^2 - (\partial_0 A_1^a)(D_1A_0)^a + \frac{1}{2}\big((D_1 A_0)^a\big)^2 + \i \bar\psi \slashed{\partial}\psi - J^{\mu,a} A^a_\mu + m \bar\psi\psi\right)\ ,
\end{align}
where $J^{\mu,a} = g \bar\psi \gamma^\mu T^a \psi$ is the
$\text{SU}(2)$ Noether current. Integrating out the non-dynamical
field $A_0^a$ in the $A_1 = 0$ gauge, we get an effective action only
in terms of the fermions:
\begin{align}
  \mc{S}_{\rm eff}[\psi]
%  &= \int \ud^2 x \left(\i \bar\psi\slashed{\partial}\psi + m\bar\psi\psi + \frac{1}{2} J^{0,a} \frac{1}{\partial_1^2} J^{0,a}\right)\ ,\nonumber\\
%  &= \int \ud^2x\,\big( \i \bar\psi\slashed{\partial}\psi + m\bar\psi\psi\big) +   \frac{1}{4}\int \ud x^0 \ud x^1 \ud y^1 J^{0,a}(x) |x^1 - y^1| J^{0,a}(y)\ ,\nonumber\\
  &= \int \ud^2x\, \big(\i \bar\psi\slashed{\partial}\psi + m \bar\psi\psi\big) +   \frac{g^2}{4}\int \ud x^0 \ud x^1 \ud y^1 (\psi^\dag T^a \psi)(x) |x^1 - y^1| (\psi^\dag T^a \psi)(y)\ ,
\end{align}
where we have used the Fourier transform of $1/\partial_1^2$ given in
\eqref{fouriersq} of Appendix \ref{conventions}. Note that the
four-fermi term is simply the pairwise Coulomb interaction between
charge densities $J^{0,a}(x)|x^1 - y^1| J^{0,a}(y)$. In the large $J$
limit, using the simplifications of Section \ref{largeJsec}, the above
action can be written as
\begin{multline}\label{0Seff}
  \mc{S}_{\rm eff}[\psi] = \int \ud^2 x\int_0^\pi\ud\theta\sin\theta\,\Big( \i \bar\psi(\theta,x) \slashed{\partial} \psi(\theta,x) + m \bar\psi(\theta,x)\psi(\theta,x) \\ + \frac{\lambda}{2}\int_0^\pi\ud\theta'\sin\theta' \cos(\theta-\theta')\big(\psi^\dag\psi\big)(\theta,x)\frac{1}{\partial_1^2}\big(\psi^\dag \psi\big)(\theta',x)\Big)\ ,
\end{multline}
where $\lambda$ is the 't Hooft coupling $\lambda = g^2 J^2$, and the
notation $(\psi^\dag\psi)(\theta,x)$ stands for
$\psi^\dag(\theta,x)\psi(\theta,x)$ with the spinor indices contracted
between $\psi^\dag$ and $\psi$. Next, we rewrite the four-fermi term
as a quadratic term in an auxiliary local bosonic variable $\rho$ by
inserting $1$ in the path integral in terms of auxiliary fields
$\sigma$ and $\rho$:
\begin{equation}\label{idenins}
  1 \sim \int [\ud\sigma][\ud\rho] \exp\left(- \int \ud^2 x \int_0^\pi \ud\theta\sin\theta\, \sigma(\theta,x)\left[\rho(\theta,x) - \big(\psi^\dag \psi\big)(\theta,x)\right]\right)\ .
\end{equation}
\footnote{Note that the usual representation of a $\delta$-function is
  $\delta(\rho - \psi^\dag\psi) = \int \ud \sigma\,\e^{\i \sigma (\rho - \psi^\dag \psi)}$ but the $\i$ in the
  exponent is missing in \eqref{idenins} since we have rotated the
  contour of $\sigma$ to be along the imaginary axis.} The effective
action involving the fermions and auxiliary fields is
\begin{multline}\label{0Seff1}
  \mc{S}_{\rm eff}[\psi,\sigma,\rho] =  \int \ud^2 x\int_0^\pi\ud\theta\sin\theta\,\Big(  \bar\psi(\theta,x) \big(\i \slashed{\partial} - \gamma^0\i\sigma(\theta,x)\big) \psi(\theta,x) + m \bar\psi(\theta,x)\psi(\theta,x) \\  + \i\sigma(\theta,x)\rho(\theta,x) + \frac{\lambda}{2}\int_0^\pi\ud\theta'\sin\theta' \cos(\theta-\theta')\rho(\theta,x)\frac{1}{\partial_1^2}\rho(\theta',x)\Big)\ .
\end{multline}
To handle the $1/\partial_1^2$ in the last term above, we further
introduce a scalar field $\phi(\theta,x)$ such that
\begin{equation}\label{currentbos}
  \rho(\theta,x) = \partial_1\phi(\theta,x)\ .
\end{equation}
\footnote{This is abelian bosonization of a single Dirac fermion in
  $1+1$ dimensions where the $\text{U}(1)$ current
  $\bar\psi \gamma^\mu \psi$ is written in terms of a scalar field
  $\phi$ as $ \varepsilon^{\mu\nu} \partial_\nu \phi$ (we omit a
  customary $\frac{1}{\sqrt{\pi}}$ that is present in usual formulas
  for bosonization). Here, we imagine that there is one such
  $\text{U}(1)$ for each $\theta$. Then $\rho$ is simply the zeroth
  component of the $\text{U}(1)$ current
  $\rho = \psi^\dag \psi = \bar\psi \gamma^0 \psi$ which is mapped to
  $\partial_1 \phi$.} The effective action then becomes
\begin{multline}\label{0Seff2}
  \mc{S}_{\rm eff}[\psi,\sigma,\phi] =  \int \ud^2 x\int_0^\pi\ud\theta\sin\theta\,\Big(  \bar\psi(\theta,x) \big(\i \slashed{\partial} - \gamma^0\i\sigma(\theta,x)\big) \psi(\theta,x) + m \bar\psi(\theta,x)\psi(\theta,x) \\  + \i \sigma(\theta,x)\partial_1\phi(\theta,x) - \frac{\lambda}{2}\int_0^\pi\ud\theta'\sin\theta' \cos(\theta-\theta')\phi(\theta,x)\phi(\theta',x)\Big)\ .
\end{multline}

\subsection{A large $J$ saddle point}

Since the effective action is a single trace over the $\theta$ labels,
the action is $\mc{O}(J)$ in the large $J$ limit. Hence, we can use
saddle-point approximation to compute the path integral. As usual, we
look for translation-invariant saddle points. This corresponds to a
constant value for the scalar field $\phi$ along the spacetime
directions ($\phi$ can still be a non-trivial function of $\theta$).

The field $\sigma$ is not a scalar since it couples to
$\psi^\dag \psi$ which is the zeroth component of a vector
current. But it may be interpreted as the zeroth component of a
fictitious $\text{U}(1)$ gauge field $\cA_\mu$, and indeed it appears
in that role in the fermion kinetic term in \eqref{0Seff2}. The
spatial component of the gauge field $\cA_1$ is zero, and this can be
thought of as a gauge choice for the fictitious gauge field. Since the
field strength $\cF_{\mu\nu} = -\i\varepsilon_{\mu\nu} \partial_1\sigma$
is a pseudoscalar in $1+1$ dimensions, we can take it to be constant
at our translation-invariant saddle point. Thus, we assume that
$\eta = \partial_1\sigma$ is a constant (along spacetime, but not
necessarily along $\theta$).

On this saddle point, the problem becomes that of fermions in a
constant (imaginary) electric field background in $1+1$
dimensions. The effective action for this is well-known: it is the
Euler-Heisenberg effective action
\cite{Heisenberg:1936nmg,Weisskopf:1936hya}, which was re-derived by
Schwinger using the proper time method \cite{Schwinger:1951nm} (also
see \cite{Itzykson:1980rh,Schwartz:2014sze} for a pedagogical
exposition). The result of integrating out the fermions in a constant
$\eta$ background is
\begin{multline}\label{ferdetconst}
  \exp\left(\int_0^\pi \ud\theta\sin\theta\,\log\det(\i \slashed\cD + m)\right)\\ = \exp\left(-\frac{\i}{2}\int \ud^d x\int_0^\pi \ud\theta\sin\theta \int_0^\infty \frac{\ud t}{(2\pi t)^{d/2}}\e^{-m^2 t} \eta \coth t\eta\right)\ ,
\end{multline}
where we have given the expression for the fermion determinant in $d$
dimensions for the purposes of regularization. We discuss the
regularization in detail in Appendix \ref{reg} and only give the
regularized final expression here. Partially integrating the
$\partial_1$ derivative in the $\sigma\partial_1\phi$ term in
\eqref{0Seff2}, the effective action for $\eta$ and $\phi$ becomes
(up to some constant terms)
\begin{align}\label{0Seff3}
  \mc{S}_{\rm eff}[\eta,\phi]
  &=  \int \ud^2 x\int_0^\pi\ud\theta\sin\theta\,\bigg(-\frac{1}{4\pi}\int_0^\infty \frac{\ud t}{t^2} \e^{-m^2 t} (t\eta(\theta) \coth t\eta(\theta) - 1)\nonumber \\ &\qquad\qquad\qquad\qquad - \i\eta(\theta)\phi(\theta) - \frac{\lambda}{2}\int_0^\pi\ud\theta'\sin\theta' \cos(\theta-\theta')\phi(\theta)\phi(\theta') \bigg)\ ,\nonumber\\
  &=  \int \ud^2 x\int_0^\pi\ud\theta\sin\theta\,\bigg( - \frac{\eta}{4\pi}\left(2 \log\Gamma\left(\frac{m^2}{2\eta}\right) - \log 2\pi + \left(1 - \frac{m^2}{\eta}\right) \log \frac{m^2}{2\eta}\right)\nonumber \\
  &\qquad\qquad\qquad\qquad- \i\eta(\theta)\phi(\theta) - \frac{\lambda}{2}\int_0^\pi\ud\theta'\sin\theta' \cos(\theta-\theta')\phi(\theta)\phi(\theta') \bigg)\ .
\end{align}
The saddle point equations are then
\begin{align}\label{saddle}
  0 &=  -\i \phi(\theta) -\frac{1}{4\pi} \int_0^\infty \frac{\ud t}{t^2} \e^{-m^2 t} \frac{\partial}{\partial\eta} (t\eta \coth t\eta) \ ,\nonumber\\
  0 &= -\i \eta(\theta) -\lambda \int_0^\pi \ud\theta'\sin\theta'\,\cos(\theta-\theta') \phi(\theta')\ .
\end{align}
We have not been able to solve the above equations explicitly. For the
next two subsections, we assume that we have a solution for
\eqref{saddle}.

\subsection{The spectral function}

Recall from \eqref{ferdetconst} that
integrating out the fermions gave the log determinant
\begin{equation}\label{detevalalt}
\int_0^\pi \ud\theta\sin\theta\,\str\, \langle x |\log(\i \slashed\cD + m)| x\rangle
   = -\frac{\i}{2}\int_0^\pi \ud\theta\sin\theta \int_0^\infty \frac{\ud t}{t (2\pi t)^{d/2}} \e^{- m^2 t}\,  t\eta \coth t\eta\ ,
\end{equation}
where $\str$ is the trace over Dirac spinor indices. Differentiating
once w.r.t.~$m$, we get
\begin{align}\label{deteval3}
  \int_0^\pi \ud\theta\sin\theta\,\str\, \langle x |\frac{-\i}{\i\slashed\cD + m}| x\rangle  &=  m \int_0^\pi \ud\theta\sin\theta\int_0^\infty \frac{\ud t}{(2\pi t)^{d/2}} \e^{- m^2 t}\, t\eta \coth t\eta\ .
\end{align}
The left-hand side is nothing but the trace (over spacetime, Dirac
indices and gauge indices) of the fermion two-point function at
coincident points $x=y$. Since we expect our saddle-point to be
unitary and translation invariant, we can write down a
K\"all\'en-Lehmann representation for the left-hand side so that
\begin{equation}\label{spectraleq}
  \int\frac{\ud^d p}{(2\pi)^d}  \int_0^\infty \ud\mu \frac{2^{d/2} \i  \varrho(\mu)}{p^2 - \mu + \i\varepsilon} =  m \int_0^\pi \ud\theta\sin\theta\int_0^\infty \frac{\ud t}{(2\pi t)^{d/2}} \e^{- m^2 t}\, t\eta \coth t\eta\ .
\end{equation}
\footnote{We write this in analogy with the case of a single free
  Dirac fermion of mass $m$ in $d$ dimensions:
  \begin{equation}
    \str\,\langle x | \frac{-\i}{\i \slashed{\partial} + m} | x\rangle = \str\,\langle x | \frac{-\i(\i\slashed{\partial} - m)}{- \slashed{\partial}^2 - m^2} | x\rangle =  \int \frac{\ud^d p}{(2\pi)^d} \frac{2^{d/2} \i m}{p^2 - m^2 + \i\varepsilon} = \int \frac{\ud^d p}{(2\pi)^d} \int_0^\infty \ud\mu\frac{2^{d/2} \i \varrho(\mu)}{p^2 - \mu + \i\varepsilon}\ ,\nonumber
  \end{equation}
  with $\varrho(\mu) = m\delta(\mu - m^2)$ for the free fermion.\label{specfoot}} By
recasting the left-hand side of the above equation \eqref{spectraleq}
into a proper-time integral, we can hope to extract $\varrho(\mu)$ which
contains spectral information of the theory. Introducing the Schwinger
proper-time integral for the denominator of the integrand, we get
\begin{equation}\label{aa}
  \int\frac{\ud^d p}{(2\pi)^d}  \int_0^\infty \ud\mu \frac{ 2^{d/2} \i \varrho(\mu)}{p^2 - \mu + \i\varepsilon} =  2^{d/2}\int\frac{\ud^d p}{(2\pi)^d}  \int_0^\infty \ud\mu\, \varrho(\mu)\int_0^\infty \ud s\, \e^{\i s (p^2 - \mu + \i\varepsilon)}\ .
\end{equation}
Performing the Gaussian integrals over the momenta % (after Wick
% rotation $p^0 = \i p^0_E$), we get
% \begin{equation}
%   \int\frac{\ud^d p}{(2\pi)^d}   \e^{\i s p^2} = \i \int\frac{\ud^d p_E}{(2\pi)^d}   \e^{-\i s p^2_E} = \frac{\i}{(4\pi \i s)^{d/2}}\ .
% \end{equation}
and the Wick rotation $s \to -\i t$, we get
\begin{equation}\label{propspectral}
   \int\frac{\ud^d p}{(2\pi)^d}  \int_0^\infty \ud\mu \frac{2^{d/2} \i \varrho(\mu)}{p^2 - \mu + \i\varepsilon} =    \int_0^\infty \frac{\ud t}{(2\pi t)^{d/2}} \int_0^\infty \ud\mu\,\e^{- t\mu} \varrho(\mu)\ .
\end{equation}
Comparing \eqref{propspectral} and \eqref{spectraleq}, we
get
\begin{equation}\label{spectralcomp}
  \int_0^\infty \ud\mu\,\e^{-t\mu} \varrho(\mu) =  m \e^{-m^2 t} \int_0^\pi\ud\theta\sin\theta\,  t\eta(\theta) \coth t\eta(\theta)\ .
\end{equation}
In the limit $\eta \to 0$, we get
\begin{equation}
  \varrho(\mu) = 2 m \delta(\mu - m^2)\ ,
\end{equation}
which is the correct spectral density for a free fermion up to a
factor of $2$ (see Footnote \ref{specfoot}). This factor of $2$ is
from the integral over $\theta$ which is in fact a sum over the number
of components $2J+1$ of the fermion in units of $1/J$ in the large $J$
limit (we refer the reader to the discussion in Section
\ref{largeJsec}). We can obtain some partial information about
$\varrho(\mu)$ by looking at the late time behaviour $t \to \infty$ of
the right hand side. When $t \to\infty$, the $\coth t\eta$ factor goes
to $1$ so that the right hand side of \eqref{spectralcomp} becomes
\begin{equation}
  \int_0^\infty \ud\mu\,\e^{-t\mu} \varrho(\mu) \approx  C m t\, \e^{-m^2 t}\ ,
\end{equation}
where the constant $C$ is
$C = \int_0^\pi\ud\theta\sin\theta\, \eta(\theta)$. The $\varrho(\mu)$
that corresponds to the above function of $t$ is
\begin{equation}\label{latetime}
  \varrho(\mu) = C m \delta'(\mu - m^2)\ .
\end{equation}
Clearly, this does not signify a simple pole in the propagator since
it is not a $\delta$-function. In fact, the full spectral density
corresponding to \eqref{spectralcomp} can be computed by an inverse
Laplace transform. We present the calculation in Appendix
\ref{spectral} and give the result here:
\begin{equation}
  \varrho(\mu) =  2m\, \Theta(\mu - m^2) \int_0^\pi \ud\theta\sin\theta\, \sum_{k=-\infty}^\infty \delta'\left(\frac{\mu - m^2 - 2k \eta(\theta)}{\eta(\theta)}\right) \ ,
\end{equation}
where the derivative on the $\delta$-function is w.r.t.~$\mu$. The
$k=0$ term indeed corresponds to the late time answer \eqref{latetime}
(after choosing $\Theta(\mu-m^2)|_{\mu = m^2} = \frac{1}{2}$).

\subsection{Majorana fermions}
For Majorana fermions, the same calculation as for the Dirac fermions
goes through, but with the density variables $\rho(\theta,x)$
satisfying a constraint that arises from the Majorana
condition. Recall that, for a Majorana fermion $\Psi(\theta,x)$, the
density variable is
$\rho(\theta,x) = \Psi^\dag(\theta,x)\Psi(\theta,x)$. Using the same
manipulations as in \eqref{contreal} in Section \ref{majorana}, we see
that $\rho(\theta,x)$ must satisfy
\begin{equation}\label{rhoiden}
  \rho(\pi-\theta,x) = -\rho(\theta,x)\ .
\end{equation}
In particular, this implies the boundary condition
$\rho(\theta=\pi/2, x) = 0$. This corresponds to the fact that the
component $\Psi_0$ corresponding to the $\text{SU}(2)$ representation
index $n=0$ of the fermion is real or imaginary (depending on $J$
being even or odd) so that $\Psi^\dag_0\Psi_0$ is identically
zero. Without loss of generality, we can assume that the field
$\sigma(\theta,x)$ also satisfies the same condition
\eqref{rhoiden}. The rest of the analysis is the same as for the Dirac
fermion in an integer spin $J$ representation.

\section{Discussion}

In this work, we have analyzed $\text{SU}(2)$ gauge theory coupled to
a single massive quark in the spin $J$ representation of the gauge
group. In a 't Hooft-like large $J$ limit where we keep
$\lambda= g^2 J^2$ fixed as $J$ is taken large, we have obtained an
integral equation for the gauge invariant meson wavefunctions. This
integral equation is the same for the two cases of Dirac fermions for
arbitrary $J$ and Majorana fermions for integer $J$. However, in the
case of Majorana fermions for integer $J$, the solutions $\Phi(x)$ of
the integral equation have to be odd under $x \to 1-x$ as a
consequence of the Majorana reality condition. This retains only half
of the spectrum of the integral equation. For the case of zero quark
mass $m=0$, this projection removes the zero-mass bound state solution
of the integral equation for meson wavefunctions, resulting in a gap
in the meson spectrum.

To the extent that we can compare with results at finite $J$ (in a
systematic expansion in $1/J$ by retaining subleading terms in the
effective action), the above results match with well-known facts about
the the $\text{SU}(2)$ theory with quarks in the adjoint
representation: (1) the model with massive quarks has a completely
discrete spectrum due to the linear confining potential
\cite{Dalley:1992yy,Kutasov:1993gq,Boorstein:1993nd,Kutasov:1994xq,
  Gross:1995bp,Gross:1997mx,Boorstein:1997hd,Cherman:2019hbq,
  Komargodski:2020mxz,Dempsey:2022uie}, and (2) the model with
massless quarks does not have a zero-mass bound state. However, it has
been argued in \cite{Delmastro:2021otj,Delmastro:2022prj} that the
$\text{SU}(2)$ theory with massless spin $J$ quarks for any $J > 2$ is
gapless: there is a massless state in the spectrum of the Hamiltonian
which is created by the operator $T - T'$ where $T$ and $T'$ are the
stress tensors of the $\text{SO}(2J+1)_1$ and the
$\text{SU}(2)_{I(J)}$ WZW models respectively, where
$I(J) = J(J+1)(2J+1)/3$ being the index of the spin $J$
representation. This is not in contradiction with our results in the
large $J$ limit since our method diagonalizes the Hamiltonian only in
the two fermion sector (in other words, the integral equation is for
the meson wavefunctions only) whereas the massless state involves
contributions from four-fermion states as well since the stress tensor
$T'$ for the $\text{SU}(2)_{I(J)}$ model involves four fermions (as is
clear from the Sugawara form of the stress tensor). It is plausible
that such massless states may indeed be observed if one is able to
extend the methods of this paper to include states with higher numbers
of fermions.

The $\text{SU}(2)$ theory coupled to an adjoint quark is known to have
a screening vs. confinement transition at zero quark mass. Does this
transition persist for larger integer $J$?  For $J=1$, the adjoint
quark is certainly not capable of screening external charges in the
fundamental representation. However, it was shown in
\cite{Gross:1995bp} that there exist certain topological excitations
at $m=0$ which transform in the fundamental representation of
$\text{SU}(2)$ and hence are capable of screening fundamental external
charges. The same was argued using mixed 't Hooft anomalies between
the 0-form $\mathbf{Z}_2$ chiral symmetry and the 1-form
$\mathbf{Z}_2$ centre symmetry \cite{Cherman:2019hbq}, and
non-invertible topological line operators for the $\text{SU}(N)$
adjoint theory \cite{Komargodski:2020mxz}. It would be interesting to
see if such an analysis can be performed for spin $J$ massless quarks
in $\text{SU}(2)_{1+1}$ QCD.

Our method of calculation can be generalized to $\text{SU}(N)$ QCD
with quarks in representations having large representation weights
$(J_1,\ldots,J_{N-1})$. Recall that the discrete representation index
$m = -J,\ldots, J$ for the spin $J$ representation was the eigenvalue
of the Cartan generator $T^3$, and it was replaced by a continuum
angle $\theta \in [0,\pi]$. This suggests that, in the large weight
limit $J_i \to \infty$ of $\text{SU}(N)$ representations, we should
associate one continuum angle for each Cartan generator of
$\text{SU}(N)$, giving a total of $N-1$ angles. It would also be
interesting to explore the double limit where $J_i \to \infty$ and
$N \to \infty$ with some ratios of $J_i$ and $N$ fixed in the limit.

\acknowledgments{S.~R.~W.~would like to thank David Gross and Luis
  Alvarez-Gaume for early discussions, and the Infosys Foundation Homi
  Bhabha Chair at ICTS-TIFR for its generous support. The authors
  would also like to thank R.~Loganayagam for discussions, and Gautam
  Mandal, Shiraz Minwalla for comments on the manuscript. This work is
  supported by the Department of Atomic Energy under project
  no. RTI4001.}

\appendix

\section{Conventions}\label{conventions}
The lightcone coordinates $x^\pm$ are related to Cartesian coordinates
as $x^{\pm} = x^0 \pm x^1$, $p_{\pm} = \frac{1}{2} (p_0 \pm p_1)$ and
so on. The Minkowski metric is
$(\ud x^{0})^2 - (\ud x^1)^2 = \ud x^+ \ud x^-$.  We also use the
$\gamma$ matrices
\begin{equation}
  \gamma^0 = \sigma_1\ ,\quad \gamma^1 = -\i \sigma_2\ ,\quad \gamma_c = \gamma^0 \gamma^1 = \sigma_3\ .
\end{equation}
We write the two-component fermion $\Psi_\alpha$ and the components
$(\gamma_\mu)_\alpha{}^\beta$ of a $\gamma$-matrix as
\begin{equation}
  \Psi_\alpha = \pmat{\Psi_- \\ \Psi_+}\ ,\quad \gamma_\mu =  \pmat{(\gamma_\mu)_-{}^- & (\gamma_\mu)_-{}^+ \\ (\gamma_\mu)_+{}^- & (\gamma_\mu)_+{}^+}\ .
\end{equation}
The charge conjugation matrix $C$ is defined via
$C\gamma^\mu C^{-1} = -\gamma^{\mu,T}$, that is, $$C^{\alpha\gamma} (\gamma^\mu)_\gamma{}^\delta (C^{-1})_{\delta\beta} = - (\gamma^{\mu,T})^\alpha{}_\beta\ .$$
It is numerically given by $C = \i\sigma_2$, that is,
\begin{equation}
  \pmat{C^{--} & C^{-+} \\ C^{+-} & C^{++}} = \pmat{0 & 1 \\ -1 & 0}\ .
\end{equation}
We use $C_{\alpha\beta}$ to raise and lower the spinor indices so that
we have $\Psi^\alpha = (C^{-1})^{\alpha\beta}\Psi_\beta$ and
$\Psi_\alpha = C_{\alpha\beta}\Psi^\beta$. %  Explicitly, we have
% \begin{equation}
% \pmat{\Psi^- \\ \Psi^+} = \pmat{0 & -1 \\ 1 & 0} \pmat{\Psi_- \\ \Psi_+} = \pmat{-\Psi_+ \\ \Psi_-}\ . 
% \end{equation}
The conjugate spinor is defined as $\bar\Psi = \Psi^\dag \gamma^0$. In
our conventions, complex conjugation changes the location of an index
so that $\Psi^\dag$ has an upper index whereas $\Psi^T$ has a lower
index.  In components, we have
$\bar\Psi{}^\beta = (\Psi^\dag)^\alpha (\gamma^0)_\alpha{}^\beta$. %  and
% \begin{equation}
%   \pmat{\bar\Psi{}^- & \bar\Psi{}^+} = \pmat{(\Psi_-)^* & (\Psi_+)^*}\pmat{0 & 1 \\ 1 & 0} = \pmat{(\Psi_+)^{*} & (\Psi_-)^{*}}\ .
% \end{equation}
We define the Majorana conjugation operation on $\Psi$ to be
$\Psi\longrightarrow \Psi^T C$. A Majorana fermion $\Psi$
satisfies $\Psi^\dag\gamma^0 = \Psi^T C$, that is,
$(\Psi^\dag)^\alpha (\gamma^0)_\alpha{}^\beta = \Psi_\alpha
C^{\alpha\beta}$:
% \begin{equation}
%   \pmat{(\Psi_-)^{*} & (\Psi_+)^{*}}\pmat{0 & 1 \\ 1 & 0} = \pmat{\Psi_- & \Psi_+} \pmat{0 & 1 \\ -1 & 0}\ ,
% \end{equation}
% that is
\begin{equation}
   \pmat{(\Psi_+)^* & (\Psi_-)^*} = \pmat{-\Psi_+ & \Psi_-}\ ,
 \end{equation}
 so that $\Psi_-$ is real and $\Psi_+$ is imaginary.

We also need the following Fourier transforms:
\begin{equation}\label{fouriersq}
  \frac{1}{\partial^2_x} \delta(x-y) =   \frac{1}{\partial^2_x}\int \frac{\ud k}{2\pi} \e^{\i k(x-y)} = \int \frac{\ud k}{2\pi} \frac{1}{-k^2}\e^{\i k(x-y)} = -\frac{1}{2\pi} \i\pi (\i(x-y)) \sgn(x-y) = \frac{1}{2}|x-y|\ .
\end{equation}
and
\begin{equation}
  \frac{1}{\partial_x} \delta(x-y) =   \frac{1}{\partial_x}\int \frac{\ud k}{2\pi} \e^{\i k(x-y)} = \int \frac{\ud k}{2\pi} \frac{1}{\i k}\e^{\i k(x-y)} = \frac{1}{2\pi\i} \i\pi  \sgn(x-y) = \frac{1}{2}\sgn(x-y)\ .
\end{equation}

\section{Representations of $\text{SU}(2)$ and Majorana fermions}\label{SU2rep}

Recall that an irreducible representation of $\text{SU}(2)$ is
labelled by a positive number $J$ which is either an integer or a
half-integer which we call the `spin'. This terminology should cause
no confusion since there is no other notion of spin in $1+1$
dimensions, e.g., from the representations of the Lorentz group. The
spin $J$ representation is simply described in the so-called $J$-$m$
basis which consists of the basis kets
\begin{equation}
  |J,m\rangle\ ,\quad m = -J, -J +1,\ldots, J-1, J\ .
\end{equation}
We write the generators of the Lie algebra of $\text{SU}(2)$ as $T^+$,
$T^-$ and $T^a$ and they satisfy the Lie algebra $[T^+, T^-] = 2 T^3$,
$[T^3, T^\pm] = \pm T^\pm$. Their matrix elements in the above basis
are
\begin{align}\label{Tcomp}
  &\langle J,n|T^+|J,m\rangle = \sqrt{(J-m)(J+m+1)}\ \delta_{n,m+1}\ ,\nonumber \\
  &\langle J,n|T^-|J,m\rangle = \sqrt{(J+m)(J-m+1)}\ \delta_{n,m-1}\ ,\nonumber\\
  &\langle J,n|T^3|J,m\rangle = m\ \delta_{n,m}\ .
\end{align}
There is a concrete realization of the above basis in terms of degree
$2J$ polynomials in one complex variable $\zeta$
\begin{equation}
  |J,m\rangle \quad \leftrightarrow\quad \zeta^{J+m}\ ,\quad m = -J,-J+1,\ldots, J-1, J\ .
\end{equation}
The generators act on the polynomials as
\begin{equation}
  T^+ = -\zeta^2 \frac{\partial}{\partial\zeta} + 2 J \zeta\ ,\quad T^- = \frac{\partial}{\partial\zeta}\ ,\quad T^3 = \zeta \frac{\partial}{\partial\zeta} - J\ .
\end{equation}
A general element of the representation is then a general degree $2J$
polynomial:
\begin{equation}\label{O2J}
  \psi(\zeta) = \sum_{m=-J}^J \psi^{(m)} \zeta^{J+m}\ .
\end{equation}
In the polynomial basis, there is a convenient definition of the
complex conjugate representation which allows us to concretely
understand the reality properties of the various representations of
$\text{SU}(2)$. The complex conjugation consists of three steps:
\begin{enumerate}
\item Complex conjugate the coefficients $\psi^{(m)}$:
  $\psi^{(m)} \to \psi^*_{(m)}$, \footnote{We have adopted the
    convention that complex conjugation naturally changes the location
    an $\text{SU}(2)$ representation index from superscript to
    subscript and vice-versa.}
\item Replace every occurrence of $\zeta$ by $-1 / \zeta$,
\item Multiply the result by $\zeta^{2J}$.
\end{enumerate}
For instance, the above steps applied to $\psi(\zeta)$ in \eqref{O2J}
gives
\begin{multline}
  \sum_{m=-J}^J \psi^{(m)} \zeta^{J+m}\quad \substack{\longrightarrow \\ 1.} \quad   \sum_{m=-J}^J \psi^*_{(m)} \zeta^{J+m}\quad \substack{\longrightarrow \\ 2.} \quad \sum_{m=-J}^J \psi^*_{(m)} (-\zeta)^{-J-m}\\  \substack{\longrightarrow \\ 3.} \quad \zeta^{2J}\sum_{m=-J}^J \psi^*_{(m)} (-\zeta)^{-J-m}\ .
\end{multline}
Note that the end result is
again a polynomial of degree $2J$ and the transformation is
\begin{equation}\label{ccrep}
  \psi(\zeta) \to \psi^*(\zeta) = \sum_{m=-J}^J (-1)^{-J-m}\psi^*_{(m)} \zeta^{J-m}\ .
\end{equation}
Let us see what happens if we apply the above operation twice:
\begin{multline}
  \sum_{m=-J}^J \psi^{(m)} \zeta^{J+m}\quad \longrightarrow\quad   \sum_{m=-J}^J (-1)^{-J-m} \psi^*_{(m)} \zeta^{J-m}\\ \longrightarrow \quad  \zeta^{2J} \sum_{m=-J}^J (-1)^{-J-m} \psi^{(m)} (-\zeta)^{-J+m}\ .
\end{multline}
The final expression $\psi^*{}^*(\zeta)$ is simply
\begin{equation}\label{doublecc}
  \psi^*{}^*(\zeta)  = (-1)^{-2J} \psi(\zeta) \ .
\end{equation}
Observe that when $J$ is half-integer (resp.~integer), the `complex
conjugation' operation squares to $-1$ (resp.$+1$).

Thus, when $J$ is an integer, we have a legitimate complex conjugation
operation which squares to $+1$. We can use this to impose a reality
condition on the representation:
\begin{equation}
  \psi^*(\zeta) = \psi(\zeta)\quad\text{for integer $J$ only.}
\end{equation}
For components, the reality condition can be obtained by comparing the same powers of $\zeta$ in \eqref{O2J} and \eqref{ccrep}:
\begin{equation}\label{realityspinJ}
  \psi^*_{(m)} = (-1)^{J+m} \psi^{(-m)}\ .
\end{equation}
Since the complex conjugation operation squares to $+1$, it has two
eigenspaces corresponding to the eigenvalues $\pm 1$. In the reality
conditions above, we have projected down to the $+1$ eigenspace. We
could also project to the $-1$ eigenspace which corresponds to
\begin{equation}\label{imaginarityspinJ}
  \psi^*_{(m)} = -(-1)^{J+m} \psi^{(-m)}\ .
\end{equation}
These two conditions \eqref{realityspinJ} and \eqref{imaginarityspinJ}
split the complex representation $\sum_m \psi^{(m)} \zeta^{J+m}$
(where $J$ is integer), into a real and an imaginary part which
transform separately as irreducible representations of $\text{SU}(2)$.

Suppose we have a Dirac spinor in $1+1$ dimensions which transforms in
the spin $J$ representation of $\text{SU}(2)$ with $J$ integer. Then,
the Majorana conjugation operation and the above complex conjugation
operation \eqref{realityspinJ} for integer $J$ can be simultaneously
applied. Thus, a Dirac spinor $\psi$ transforming in the spin $J$
representation of $\text{SU}(2)$ with $J$ integer splits into two
Majorana spinors $\Psi_1$ and $\Psi_2$ transforming in the spin $J$
(integer) representation. % , and the customary normalization relating the
% two is
% \begin{equation}\label{diractomaj}
%   \psi = \frac{\Psi_1 + \i \Psi_2}{\sqrt{2}}\ .
% \end{equation}
The Majorana condition in our choice of $\gamma$-matrices for an
$\text{SU}(2)$ inert Dirac spinor $\Psi$ is (see Appendix
\ref{conventions})
\begin{equation}
  \Psi_-^* = \Psi_-\ ,\quad \Psi^*_+ = - \Psi_+\ ,
\end{equation}
where $\Psi_\pm$ are the spinor components of $\Psi$. For a Dirac
spinor in a spin $J$ representation with $J$ integer, we combine the
Majorana condition with the reality condition \eqref{realityspinJ} to
get
\begin{equation}\label{realityJm}
  \Psi^*_{m,-}(x) = (-1)^{J+m} \Psi^{-m}_-(x)\ ,\quad  \Psi^*_{m,+}(x) = -(-1)^{J+m} \Psi^{-m}_+(x)\ .
\end{equation}
We have dropped the parentheses around the indices $m$ in
$\Psi^m_{\pm}$ since there is no danger of confusing them with
exponents as in $\zeta^{J+m}$.

We make one final comment about half-integer $J$
representations. Since the `complex conjugation' operation
\eqref{ccrep} squares to $-1$ in this case (see \eqref{doublecc}),
imposing a reality condition like \eqref{realityspinJ} kills the
representation. However, if we have two instances of a half-integer
spin representation, then a reality condition mixing the two can be
imposed. Suppose the representations are $\Psi^{Am}$ with
$A=1,2$. Then, the new reality condition uses the totally
antisymmetric tensor on two indices $\varepsilon_{AB}$:
\begin{equation}\label{newccrep}
  \Psi^*_{Am} = \varepsilon_{AB} (-1)^{J+m} \Psi^{B,-m}\ .
\end{equation}
It is easy to see that the above operation squares to identity rather
than minus identity:
\begin{multline}
\Psi^*{}^*{}^{Am} = \varepsilon^{AB} (-1)^{J+m} \bar\Psi_{B,-m} = \varepsilon^{AB} (-1)^{J+m} \varepsilon_{BC} (-1)^{J-m} \Psi^{Cm} \\ = -\delta^A_C (-1)^{2J} \Psi^{Cm} = \Psi^{Am}\ ,\nonumber
\end{multline}
where we have used $\varepsilon^{AB}\varepsilon_{BC} =
-\delta^A_C$. The new reality condition can be adopted for a pair of
Dirac spinors as
\begin{equation}\label{halfrealityJm}
  \Psi^*_{Am,-}(x) = (-1)^{J+m}\varepsilon_{AB} \Psi^{B,-m}_-(x)\ ,\quad  \Psi^*_{Am,+}(x) = -(-1)^{J+m}\varepsilon_{AB} \Psi^{B,-m}_+(x)\ .
\end{equation}

In this paper, we work with a single fermion in the spin $J$
representation of the $\text{SU}(2)$ gauge group. When $J$ is an
integer, we can take the fermion to be either a Majorana spinor or a
Dirac spinor. When $J$ is a half-integer, we must take the fermion to
be a complex Dirac spinor since the new reality condition
\eqref{newccrep} only applies to two fermions (or, in general, any
even number of fermions).

\section{Coadjoint orbits of $W_\infty^{2J+1}$}\label{coadj}

In this section, we give a brief overview of the coadjoint orbit
method of Kirillov \cite{Kirillov1968,Kirillov:1976} as applied to the
$W_\infty^{2J+1} = W_{\infty} \otimes \text{U}(2J+1)$ algebra
\cite{Dhar:1994ib,Das:1991uta,Dhar:1992hr,Dhar:1992rs,Dhar:1993jc}. We
set $N = 2J+1$ for convenience. The algebra $W_\infty^{N}$ can be
described by generators
$\mg(\alpha,\beta)_{ij} = \mg(\alpha,\beta) \otimes E_{ij}$ with
\begin{equation}\label{Winfgen}
  \mg(\alpha,\beta) = \e^{\i(\alpha \bx - \beta \bp)}\ ,\quad (E_{ij})^k_l = \delta_i^k \delta_{jl}\ ,
\end{equation}
where $\bx$ and $\bp$ are the position and momentum operators on the
single particle Hilbert space and $E_{ij}$ are the elementary matrices
which generate the Lie algebra of $\text{U}(N)$. Note that the
$\mg(\alpha,\beta)$ are elements of the Heisenberg-Weyl group and
satisfy the group multiplication
\begin{equation}
  \mg(\alpha,\beta) \mg(\alpha',\beta') = \e^{\i(\alpha\beta' - \alpha'\beta)/2} \mg(\alpha+\alpha',\beta+\beta')\ .
\end{equation}
The generators \eqref{Winfgen} satisfy the commutation relations which
define the $W_\infty^N$ algebra:
\begin{multline}
  [\mg(\alpha,\beta)_{ij} , \mg(\alpha',\beta')_{kl}]
  \\ = \e^{\i(\alpha\beta' - \alpha'\beta)/2} \delta_{jk}\,
  \mg(\alpha+\alpha',\beta+\beta')_{il} -
  \e^{-\i(\alpha\beta' - \alpha'\beta)/2} \delta_{il}\,
  \mg(\alpha+\alpha',\beta+\beta')_{kj}\ .
\end{multline}
An element $\bm\Theta$ of the $W_\infty^{N}$ algebra can be expanded
in the above basis as
\begin{equation}
  \bm\Theta = \int \ud\alpha\ud\beta\,\theta(\alpha,\beta)_{ij} \mg(\alpha,\beta)_{ij}\ .
\end{equation}
For our current situation, we take the dual space $\Gamma$ of the
$W_\infty^N$ algebra to be a copy of the underlying vector space of
the $W_\infty^{N}$ algebra itself. Consequently, an element $\bm\phi$
of $\Gamma$ can also be expanded in the above basis of generators as
\begin{equation}
  \bm\phi = \int\ud\alpha \ud\beta\, \phi(\alpha,\beta)_{ij} \mg(\alpha,\beta)_{ij}\ .
\end{equation}
The pairing between the Lie algebra and its dual space is then
\begin{equation}\label{trWinf}
\tr(\bm\phi \bm\Theta) =  \int\ud\alpha\ud\beta\, \theta(\alpha,\beta)_{ij} \phi(\alpha,\beta)_{ji}\ .
\end{equation}
The adjoint action of the $W_\infty^N$ algebra is given by the
commutator:
\begin{equation}
  \delta_{\bm\eta} \bm\Theta = \i[\bm\eta, \bm\Theta]\ .
\end{equation}
\footnote{Note that the $\text{U}(1)$ subalgebra generated by
  $\bm{1} = \mg(0,0) \otimes \bm{1}_{N \times N}$ acts trivially in
  the adjoint action.} The coadjoint action is then inferred from the
inner product \eqref{trWinf} as
\begin{equation}
  \delta_{\bm\eta} \bm\phi = - \i[\bm\eta, \bm\phi]\ .
\end{equation}
According to the theory of quantization of coadjoint orbits
\cite{Kirillov:1976,Kostant,Souriau}, representations of the
$W_\infty^N$ algebra can be obtained from appropriate orbits in the
dual space $\Gamma$ of the coadjoint action of $W_\infty^N$. One
interprets a coadjoint orbit as the classical phase space of a
dynamical system and supplies a symplectic form on the phase space
that arises from the Lie bracket of the Lie algebra. One then
quantizes the phase space by any method that is admissible in the
context. In the current paper, we employ path integral quantization of
coadjoint orbits \cite{Alekseev:1988tj,Alekseev:1988vx}. The
application of the path integral method to the current context was
worked out in detail in
\cite{Dhar:1994ib,Das:1991uta,Dhar:1992hr,Dhar:1992rs,Dhar:1993jc}.

A typical coadjoint orbit of a Lie algebra is specified by the values
of the Casimir operators of the Lie algebra which are invariant under
the coadjoint action. In the present problem, the information about
all the Casimirs can be fixed in terms of two constraints involving
the coadjoint field $\bm\phi$:
\begin{equation}\label{fercoadj}
  \bm\phi^2 = \bm\phi\ ,\quad \tr(\bm{1} - \bm\phi) = Q\ ,
\end{equation}
where $Q$ is the conserved charge corresponding to the
$\text{U}(1) \subset \text{U}(N)$ symmetry. If the entire
$\text{U}(N)$ group arises as a gauge invariance of a model, then we
must set $Q = 0$ to focus on gauge invariant configurations. If the
gauge invariance is $\text{SU}(N)$, then $Q$ is the baryon number:
different choices of $Q$ corresponds to sectors with different baryon
number. In the current paper, we look at the sector with no baryons
i.e.~$Q = 0$.

The path integral quantization of the above coadjoint orbit proceeds
as follows. The symplectic form $\Omega$ at the point $\bm\phi$ on the
coadjoint orbit is described by its contraction with two tangent
vectors $\delta_1\bm\phi$, $\delta_2\bm\phi$ at the point $\bm\phi$:
\begin{equation}
  \Omega_{\bm\phi}(\delta_1\bm\phi, \delta_2\bm\phi) = \i\tr(\bm\phi [\bm\ell_1,\bm\ell_2])\ ,
\end{equation}
where $\bm\ell_1$ and $\bm\ell_2$ are the $W_\infty^N$ algebra
elements whose action on the coadjoint orbit generate the tangent
vectors $\delta_1\bm\phi$ and $\delta_2\bm\phi$ respectively. The
action can then be written as follows. Let us consider a two
dimensional region spanned by the time variable $t$ (which is $x^+$ in
the current situation) and an auxiliary variable $s$ and let us take the field $\bm\phi$ to be a function of both $t$ and $s$ with the boundary condition
\begin{equation}
  \bm\phi(s=0,t) = \bm\phi(t)\ ,\quad \lim_{s \to -\infty} \bm\phi(s,t) = \text{constant}\ .
\end{equation} 
Then we can take the two tangent vectors at $\bm\phi$ to be
$\partial_t \bm\phi$ and $\partial_s \bm\phi$. The corresponding
$\text{W}_\infty^N$ elements $\bm\ell_t$ and $\bm\ell_s$ are defined as
\begin{equation}\label{elldef}
  \partial_t \bm\phi = -\i[\bm\ell_t, \bm\phi]\ ,\quad   \partial_s \bm\phi = -\i[\bm\ell_s, \bm\phi]\ .
\end{equation}
It is easy to check that the following satisfy the above equations by
repeatedly using the defining equation $\bm\phi^2 = \bm\phi$ of the
coadjoint orbit and its derivatives w.r.t.~$t$ and $s$:
\begin{equation}\label{ellsol}
  \bm\ell_t = \i [\partial_t \bm\phi, \bm\phi]\ ,\quad   \bm\ell_s = \i [\partial_s \bm\phi, \bm\phi]\ .
\end{equation}
The action for the coadjoint orbit is then
\begin{equation}\label{Sphi}
  \mc{S}[\bm\phi] = \i\int \ud t \ud s\,\tr(\bm\phi[\bm\ell_t,\bm\ell_s]) - \int \ud t H(\bm\phi) = \i\int \ud t\ud s\,\tr(\bm\phi[\partial_t\bm\phi, \partial_s\bm\phi]) - \int \ud t H(\bm\phi)\ ,
\end{equation}
where we have included a Hamiltonian term as well; the last equality
is again achieved by plugging in \eqref{ellsol} and using
$\bm\phi^2 = \bm\phi$ repeatedly.\footnote{Compare this with the
  standard way in which the action $\int_{\mc{C}} p \ud q$ arises in
  classical mechanics from the integral of the symplectic form over a
  two dimensional region $\Sigma$ in phase space with
  $\mc{C} = \partial\Sigma$ being the boundary of $\Sigma$:
\begin{equation}
  \int_\Sigma \ud p \wedge \ud q = \int_\Sigma \ud (p \ud q) = \int_{\mc{C}} p \ud q\ .\notag
\end{equation}}

We now make contact with the problem considered in this
paper. Consider the bilocal operator
$M(x,y)_{ij} = \psi^i(x)\psi^*_j(y)$ defined in the main text
\eqref{Mdef} (for the sake of brevity, we replace the continuum
$\text{U}(2J+1)$ labels $\theta$,$\theta'$ by their discrete
counterparts $i$,$j$, and suppress the dependence on time $x^+$, and the minus
signs on $x^-$ and $y^-$). Write this as the matrix element of an
operator $\mM$ between two states $|x\rangle$, $|y\rangle$ in the
single particle Hilbert space, $M(x,y) = \langle x | \mM | y
\rangle$. Then, $\mM$ can be thought of as an element of the
$W_\infty^N$ algebra (recall $N = 2J+1$) and can be decomposed in the
$\mg(\alpha,\beta)_{ij}$ basis as
\begin{equation}\label{Mdecomp}
  \mM = \int\ud\alpha\ud\beta\, W(\alpha,\beta)_{ij} \mg(\alpha,\beta)_{ij}\ ,
\end{equation}
where
\begin{equation}\label{Wpsirel}
  W(\alpha,\beta)_{ij} = \int \ud x\, M(x+\tfrac{1}{2}\beta, x-\tfrac{1}{2}\beta) \e^{\i\alpha x}\ .
\end{equation}
The above relations \eqref{Mdecomp},\eqref{Wpsirel} follow from the
formula
\begin{equation}
  \langle x| \mg(\alpha,\beta) |y \rangle = \delta(x - y + \beta) \exp\left(\i\alpha\frac{x+y}{2}\right)\ .
\end{equation}
Recall the commutation relations of the $M(x,y)_{ij}$ from the main
text \eqref{bial}:
\begin{equation}\label{bialapp}
  [M(x_1,y_1)_{ij},M(x_2,y_2)_{kl}] = \delta_{jk} \delta(y_1 - x_2) M(x_1,y_2)_{il} 
  - \delta_{il} \delta(x_1^{-} - y_2^{-}) M(x_2,y_1)_{kj}\ .
\end{equation}
It is then straightforward to check that the $W(\alpha,\beta)_{ij}$ satisfy the defining commutation relations of the $W_\infty^N$ algebra:
\begin{multline}
  [W(\alpha,\beta)_{ij} , W(\alpha',\beta')_{kl}]
  \\ = \e^{\i(\alpha\beta' - \alpha'\beta)/2} \delta_{jk}\,
  W(\alpha+\alpha',\beta+\beta')_{il} -
  \e^{-\i(\alpha\beta' - \alpha'\beta)/2} \delta_{il}\,
  W(\alpha+\alpha',\beta+\beta')_{kj}\ .
\end{multline}
Recall from the main text that the bilocal operators satisfy the
constraint \eqref{idempot}:
\begin{equation}\label{idempotapp}
 \sum_k \int \ud z M(x,z)_{ik} M(z,y)_{kj} = M(x,y)_{ij} \big(1 + Q\big)\ ,\quad\text{i.e.}\quad \mM^2 = \mM (1+Q)\ .
\end{equation}
where $Q$ is the conserved global $\text{U}(1)$ charge operator
(baryon number operator) \eqref{conscharge}:
\begin{equation}\label{conschargeapp}
  Q = \int \ud x \sum_m \psi^*_{m-}(x^+,x^-) \psi^m_-(x^+,x^-) = \tr(\bm{1} - \bm{M})\ .
\end{equation}
We can take the expectation value of the above equations in states of
the fermion Hilbert space corresponding to zero baryon number $Q =
0$. These states form an orbit under the action of the group
$GW_\infty^N$ corresponding the $W_\infty^N$ algebra, which consists
of the elements
\begin{multline}
  |\bm\phi\rangle =  \exp\left(\i \int\ud\alpha\ud\beta\, \phi(\alpha,\beta)_{ij} W(\alpha,\beta)_{ij}\right) |0\rangle\ ,\\ \text{for all sufficiently well-behaved functions $\phi(\alpha,\beta)$},
\end{multline}
where $|0\rangle$ is the fermionic vacuum. The above are the
generalized coherent states according to Perelomov
\cite{perelomov1977} and they are elements of the coset space
$GW_\infty^N / H$ where $H$ is the subgroup that preserves $|0\rangle$
\cite{Dhar:1993jc}. But coadjoint orbits are also precisely of the
form $GW_\infty^N / H$, and thus, the states $|\bm\phi\rangle$ are in
one-to-one correspondence with points on a coadjoint orbit. Clearly,
the expectation value
$\mM_\phi = \langle \bm\phi| \mM |\bm\phi\rangle$ of the fermi bilocal
operator $\mM$ in these coherent states $|\bm\phi\rangle$ satisfy the
constraints
\begin{equation}
  \mM^2_\phi = \mM_\phi\ ,\quad \tr(\bm{1} - \mM_\phi) = 0\ ,
\end{equation}
where we have used the factorization property of coherent state
expectation values in the classical limit
$\langle \bm\phi| \mM^2 |\bm\phi\rangle = (\mM_\phi)^2 +
\mc{O}(\hbar)$ which becomes exact in the limit $\hbar \to 0$. These
equations precisely coincide with the coadjoint orbit of $W_\infty^N$
considered previously in \eqref{fercoadj} and consequently, the
classical phase space of the fermionic field theory is the coadjoint
orbit \eqref{fercoadj} of the $W_\infty^N$ algebra. Thus, the above
procedure results in a bosonization of the fermionic theory in terms
of the bosonic bilocal field $\mM_\phi$. Further, by virtue of the
$\mM_\phi$ field being an element of a $W_\infty^N$ coadjoint orbit,
there is also a ready-made action, viz., \eqref{Sphi}, that governs
its dynamics. This is the starting point of Section \ref{Mact}. In the
main text, we do not display the subscript ${}_\phi$ on $\mM_\phi$ for
the sake of brevity.

\section{Derivation of the action for the fluctuations
  $W$}\label{adjorbsim}

We reproduce the action \eqref{Maction} for the classical field
${\mM}$ here for convenience:
\begin{equation}\label{Mactionapp}
  \mc{S}[{\mM}] = \i \int_\Sigma \ud s \ud x^+ \str ({\mM} [\partial_+ {\mM}, \partial_s {\mM}]) - \int \ud x^+ \str\left(\frac{\i m^2}{8} \mS {\mM} + \frac{\lambda}{8} {\mM}\, \wt{{\mM}}\right)\ ,
\end{equation}
where the $\str$ is over the $\theta$ and $x^-$ space labels. The
fluctuations about the saddle point solution ${M}_0$ are described
by the momentum space expression \eqref{flucM}:
\begin{multline}\label{WMflucapp}
  {M}(k,k';\theta,\theta') = 2\pi\Theta(k_-)\delta(k_- - k'_-) \frac{\delta(\theta-\theta')}{\sin\theta} + \frac{\i}{\sqrt{J}} \Big(\Theta(k'_-) - \Theta(k_-)\Big) W(k,k';\theta,\theta') \\ - \frac{1}{2J} \int \frac{\ud p}{2\pi} \int_\varphi \Big(\Theta(k') + \Theta(k) - 2\Theta(p)\Big) W(k,p;\theta,\varphi) W(p,k';\varphi,\theta') + \mc{O}(J^{-3/2}) \ .
\end{multline}
Writing ${\mM} = {\mM_0} + \delta \mM$ with $\delta
{\mM}$ containing the
$W$-dependent terms in \eqref{WMfluc}, the kinetic term in
\eqref{Maction} becomes
\begin{align}
  \mc{S}_{\rm kin}[\mM] &= \i \int_\Sigma \ud s\, \ud x^+\,\str\left( {\mM}_0 [\partial_+ \delta \mM, \partial_s \delta \mM]\right)\ ,\nonumber\\
              &= \i \int_\Sigma \ud s\, \ud x^+ \str\left(\partial_+ (\mM_0 \delta \mM \partial_s \delta \mM) - \partial_s (\mM_0 \delta \mM \partial_+ \delta \mM)\right)\ .
\end{align}
We drop the $\partial_+$ term since the fluctuation is assumed to die
away at temporal infinity. Then, we get
\begin{align}
  \mc{S}_{\rm kin}[\mM] &= -\i \int \ud x^+ \int_{k,p,q}\left({M}_0(k,p) \delta {M}(p,q) \partial_+ \delta {M}(q,k)\right)\ ,\nonumber\\
              % &= \i \frac{1}{J} \int \ud x^+ \int_{k,p,q} \str\left(2\pi \Theta(k)\delta(k-p) W(p,q) (\Theta(q) - \Theta(p))  \partial_+ W(q,k) (\Theta(k) - \Theta(q))\right)\ ,\nonumber\\
              % &= \i \frac{1}{J} \int \ud x^+ \int_{p,q} \Theta(p)  (\Theta(q) - \Theta(p))  (\Theta(p) - \Theta(q))\str\left( W(p,q)  \partial_+ W(q,p)\right)\ ,\nonumber\\
              &= \i \frac{1}{J} \int \ud x^+ \int_{p>0} \int_q  (\Theta(q) - 1)  ( 1 - \Theta(q))\str\left( W(p,q)  \partial_+ W(q,p)\right)\ ,\nonumber\\
              &= -\i \frac{1}{J} \int \ud x^+ \int_{p>0} \int_{q<0}  \str\left( W(p,q)  \partial_+ W(q,p)\right)\ .
\end{align}
where an integral with a subscript like $\int_p$ without an explicitly
displayed measure is shorthand for
$\int_{-\infty}^\infty \frac{\ud p}{2\pi}$. Finally, relabelling
$q \to -q$, we get
\begin{align}\label{kin}
  -\i \frac{1}{J}  \int \ud x^+ \int_{p>0} \int_{q>0}  \str\left( W(p,-q)  \partial_+ W(-q,p)\right)\ .
\end{align}
Using the notation,
\begin{align}
  &W^{\rm pp}(q,r) = W(q,r)\ ,\quad W^{\rm pn}(q,r) = W(q,-r)\ ,\nonumber\\
  &W^{\rm np}(q,r) = W(-q,r)\ ,\quad W^{\rm nn}(q,r) = W(-q,-r)\ ,
\end{align}
the kinetic term can be written as
\begin{align}\label{kin1}
  -\i \frac{1}{J}  \int \ud x^+ \int_{p>0} \int_{q>0}  \str\left( W^{\rm pn}(p,q)  \partial_+ W^{\rm np}(q,p)\right)\ .
\end{align}
The Hamiltonian part of the action in momentum space is
\begin{align}
-  \frac{\i m^2}{4}  \int \frac{\ud k}{2\pi} \frac{1}{-\i k} \str\, {M}(k,k) - \frac{\lambda}{4} \int_{p,q,k} \frac{1}{-k^2}\, \str\big( {M}(p,q)  \wt{{M}}(q-k,p-k)\big)\ ,
\end{align}
where the traces are in the $\theta$ space and
$\wt{{M}}(q-k,p-k;\theta,\theta') = {M}(q-k,p-k;\theta,\theta')
\cos(\theta-\theta')$. The mass term gets the leading contribution
from the $\frac{1}{J} WW$ term in \eqref{WMfluc} because the
$\frac{1}{\sqrt{J}} W$ term evaluates to zero. Thus,
\begin{align}\label{mass}
  - \frac{\i m^2}{4} \int \frac{\ud k}{2\pi} \frac{1}{-\i k} \str\, {M}(k,k)
  &= - \frac{\i m^2}{4} \frac{1}{J} \int_{k>0} \int_{p>0} \left(\frac{1}{\i k} + \frac{1}{\i p}\right) \str\, W(k,-p) W(-p,k)\ ,\nonumber\\
  &= -\frac{m^2}{4} \frac{1}{J} \int_{k>0} \int_{p>0} \left(\frac{1}{k} + \frac{1}{p}\right) \str\, W(k,-p) W(-p,k)\ ,\nonumber\\
    &= -\frac{m^2}{4} \frac{1}{J} \int_{k>0} \int_{p>0} \left(\frac{1}{k} + \frac{1}{p}\right) \str\, W^{\rm pn}(k,p) W^{\rm np}(p,k)\ .
\end{align}
The interaction term has three contributions at order $\frac{1}{J}$,
schematically of the form
\begin{equation}\label{intfluc}
\text{I}:\ \  {M}_0  \frac{1}{2J} \wt{WW}\ ,\quad \text{II}:\ \ \frac{1}{2J} WW  \wt{{M}}_0\ ,\quad \text{III}:\ \ \frac{1}{\sqrt{J}} W  \frac{1}{\sqrt{J}} \wt{W}\ ,
\end{equation}
where $\wt{X}(\theta,\theta')$ corresponding to any bilocal field
$X(\theta,\theta')$ is
$\wt{X}(\theta,\theta') = \cos(\theta-\theta')X(\theta,\theta')$.  The
first two terms are equal and combine to give
\begin{multline}\label{IandII}
  \text{I} + \text{II} = -\frac{1}{J} \int_{\theta,\theta',\varphi} \int_k \frac{1}{-k^2} \int_{p,q,r} \Big( \Theta(p) + \Theta(q) - 2\Theta(r)\Big) \\{M}_0(q-k,p-k;\theta,\theta') \cos(\theta-\theta')W(p,r;\theta',\varphi) W(r,q;\varphi,\theta)\ ,
\end{multline}
Plugging in
${M}_0(q-k,p-k;\theta,\theta') = 2\pi \Theta(q-k)
\delta(q-p)\frac{\delta(\theta-\theta')}{\sin\theta}$, we get
\begin{align}
  \text{I} + \text{II} &= -\frac{1}{J} \int_{\theta,\varphi} \int_k \frac{1}{-k^2} \int_{p,r} \Big(2 \Theta(p) - 2\Theta(r)\Big) \Theta(p-k) W(p,r;\theta,\varphi) W(r,p;\varphi,\theta)\ ,\nonumber\\
                       &= -\frac{1}{J} \int_{\theta,\varphi} \int_k \frac{1}{-k^2} \int_{p,r} \Big(2 \Theta(p) - 2\Theta(r)\Big) \Theta(p-k)\,\str\big(W(p,r) W(r,p)\big)\ ,
\end{align}
Applying the $\Theta$ functions, we get
\begin{align}\label{IandIIsimpl}
\text{I} + \text{II}  &= -\frac{1}{J}\bigg[ \int_{p< 0}\int_{r>0}\int^p_{-\infty} \frac{\ud k}{-2\pi  k^2} (-2) \str\big(W(p,r) W(r,p)\big)\nonumber\\
  &\qquad\qquad+ \int_{p > 0}\int_{r < 0}\int^p_{-\infty} \frac{\ud k}{- 2\pi  k^2} (+2) \str\big(W(p,r) W(r,p)\big)\bigg]\ ,\nonumber\\
&= -\frac{1}{J} \int_{p > 0}\int_{r>0}\bigg[ \int^{-p}_{-\infty} \frac{\ud k}{-2\pi k^2} (-2) \str\big(W(-p,r) W(r,-p) \big)\nonumber\\
  &\qquad\qquad\qquad\qquad+ \int^p_{-\infty} \frac{\ud k}{-2\pi k^2} (+2) \str\big(W(p,-r) W(-r,p)\big)\bigg]\ .
\end{align} 
Using
\begin{equation}\label{identityft}
  \int_{-\infty}^\infty \frac{\ud k}{-2\pi k^2} \e^{-\i k (x-y)} = \frac{1}{2}|x-y|\quad \Rightarrow   \int_{-\infty}^\infty \frac{\ud k}{-2\pi k^2} = 0\ ,
\end{equation}
we have
\begin{equation}
  \int^p_{-\infty} \frac{\ud k}{-2\pi k^2} = - \int_{p}^{\infty} \frac{\ud k}{-2\pi k^2}\ ,
\end{equation}
and
\begin{equation}
   \left(\int^{-p}_{-\infty} + \int_r^\infty\right) \frac{\ud k}{-2\pi k^2} =    \left(\int^{-p}_{-\infty} + \int_{-p}^r + \int_r^\infty - \int_{-p}^r\right) \frac{\ud k}{-2\pi k^2} = -\int_{-p}^r \frac{\ud k}{-2\pi k^2}\ .
\end{equation}
Then, \eqref{IandIIsimpl} becomes
\begin{align}
  \text{I} + \text{II}
  &= -\frac{1}{ J} \int_{p > 0}\int_{r>0}\bigg[ \int^{-p}_{-\infty} \frac{\ud k}{-2\pi k^2} (-2) \str\big(W(-p,r) W(r,-p)\big)\nonumber\\
  &\qquad\qquad\qquad\qquad- \int_r^{\infty} \frac{\ud k}{-2\pi k^2} (+2) \str\big(W(r,-p) W(-p,r)\big)\bigg]\ ,\nonumber\\
  &= -\frac{2}{ J} \int_{p > 0}\int_{r>0} \int^{r}_{-p} \frac{\ud k}{-2\pi k^2} \,\str\big(W(-p,r) W(r,-p)\big)\ ,\nonumber\\
    &= -\frac{2}{ J} \int_{p > 0}\int_{r>0} \int^{r}_{-p} \frac{\ud k}{-2\pi k^2} \,\str\big(W^{\rm np}(p,r) W^{\rm pn}(r,p)\big)\ .
\end{align}

Next, let us look at the third term III in \eqref{intfluc}:
\begin{align}\label{III}
%  \text{III}% &= -\frac{1}{J} \int_{\theta,\theta'}\int_k \frac{1}{-k^2} \int_{p,r} \Big(\Theta(r) - \Theta(p)\Big) \Big(\Theta(p-k) - \Theta(r-k)\Big)\nonumber\\ &\qquad\qquad\qquad\qquad\qquad\qquad\qquad\qquad  W(r-k,p-k;\theta,\theta') \cos(\theta-\theta') W(p,r;\theta',\theta)\ ,\nonumber\\
             & -\frac{1}{J} \int_{\theta,\theta'}\int_k \frac{1}{-k^2} \int_{p,r} \Big(\Theta(r) - \Theta(p)\Big) \Big(\Theta(p-k) - \Theta(r-k)\Big)\,  \str\big(W(r-k,p-k)  \wt{W}(p,r)\big)\ ,
\end{align}
where recall that the $\str$ above is over the $\theta$ space and
$\wt{W}(p,r;\theta,\theta') =
\cos(\theta-\theta')W(p,r;\theta,\theta')$. Let us evaluate the
integrals. Splitting the ranges of the $p$ and $r$ integrals, we get
\begin{align}\label{III1}
  & -\frac{1}{J} \bigg[\int_{p<0} \int_{r>0} \int_{p}^r \frac{\ud k}{-2\pi k^2} (-1) \str\big( W(r-k,p-k) \wt{W}(p,r)\big)\nonumber\\
  &\qquad\qquad- \int_{p>0} \int_{r<0} \int_r^p\frac{\ud k}{-2\pi k^2} (+1) \str\big( W(r-k,p-k) \wt{W}(p,r)\big)\bigg]\ ,\nonumber\\
  &= \frac{1}{J} \int_{p>0} \int_{r>0} \int_{-p}^r \frac{\ud k}{-2\pi k^2}  \bigg[ \str\big( W(r-k,-p-k) \wt{W}(-p,r)\big) \nonumber\\
  &\qquad\qquad\qquad\qquad\qquad\qquad\qquad + \str\big( W(-p-k,r-k) \wt{W}(r,-p)\big)\bigg]\ ,\nonumber\\
  &= \frac{1}{J} \int_{p>0} \int_{r>0} \int_{-p}^r \frac{\ud k}{-2\pi k^2}  \bigg[ \str\big( W^{\rm pn}(r-k,p+k) \wt{W}{}^{\rm np}(p,r)\big) \nonumber\\
  &\qquad\qquad\qquad\qquad\qquad\qquad\qquad + \str\big( W^{\rm np}(p+k,r-k) \wt{W}{}^{\rm pn}(r,p)\big)\bigg]\ .
\end{align}
The final expressions for the interaction term becomes
\begin{align}\label{intfin}
  &-\frac{\lambda}{4}\left(\text{I} + \text{II} + \text{III}\right)\nonumber\\
  &= -\frac{\lambda}{4J} \int_{p>0} \int_{r>0} \int_{-p}^r \frac{\ud k}{-2\pi k^2}  \bigg[ \str\left( \left(\wt{W}^{\rm pn}(r-k,p+k) - W^{\rm pn}(r,p)\right)\, W^{\rm np}(p,r)\right) \nonumber\\
  &\qquad\qquad\qquad\qquad\qquad\qquad\qquad+ \str\left( \left(\wt{W}{}^{\rm np}(p+k,r-k) - W^{\rm np}(p,r)\right) W^{\rm pn}(r,p)\right)\bigg]\ .
\end{align}
The kinetic term \eqref{kin1} + mass term \eqref{mass} becomes
\begin{equation}\label{kinfin}
 \frac{1}{J}  \int_{p > 0} \int_{r > 0} \left[-\i\,\str\big(W^{\rm pn} (p,r)\partial_+ W^{\rm np}(r,p)\big) - \frac{m^2}{4}\left(\frac{1}{r} + \frac{1}{p}\right) \str\big(W^{\rm pn}(p,r) W^{\rm np}(r,p)\big)\right]\ .
\end{equation}

Varying the total action \eqref{intfin} $+$ \eqref{kinfin}
w.r.t.~$W^{\rm pn}(s,t)$, we get the following equation of motion for
$W^{\rm  np}(s,t)$:
\begin{equation}\label{Weomapp}
  -\i \partial_+ W^{\rm np}(t,s) - \frac{m^2}{4} \left(\frac{1}{s} + \frac{1}{t}\right) W^{\rm np}(t,s) 
  -\frac{\lambda}{4\pi} \int_{-t}^s \frac{\ud k}{-k^2} \Big(\wt{W}{}^{\rm np}(t+k,s-k) - W^{\rm np}(t,s)\Big) = 0\ .
\end{equation}
Displaying the $\theta$ labels explicitly, we get
\begin{multline}\label{Weomexplapp}
  -\i \partial_+ W^{\rm np}(t,s;\theta,\theta') - \frac{m^2}{4} \left(\frac{1}{s} + \frac{1}{t}\right) W^{\rm np}(t,s;\theta,\theta') \\
  -\frac{\lambda}{4\pi} \int_{-t}^s \frac{\ud k}{-k^2} \Big(W^{\rm np}(t+k,s-k;\theta,\theta')\cos(\theta-\theta') - W^{\rm np}(t,s;\theta,\theta')\Big) = 0\ .
\end{multline}

\section{Boundary condition analysis for the 't Hooft integral equation}\label{boundary}

Let us look at the principal value integral
\begin{equation}
\chi(x) =  \int_a^b \frac{\ud y}{y-x} \Phi(y)\ .
\end{equation}
Suppose the function $\Phi(x)$ has the following form near the
end-points $c=a,b$ of the contour of integration:
\begin{equation}
  \Phi(x) \sim \frac{1}{(x-c)^{\gamma_c}}\ ,\quad \gamma_c = \alpha_c + \i \alpha'_c\ ,\quad\text{with}\quad 0 \leq \alpha_c < 1\ .
\end{equation}
Then, for $x$ near $c$, we have
\begin{equation}
  \chi(x) \sim \pm \pi \cot \pi\gamma_c \frac{1}{(x-c)^{\gamma_c}}\ ,
\end{equation}
where the $+$ sign is when $c = a$ and the $-$ sign is when $c = b$
(for instance, see \cite[Chapter 4, eq.~(29.8)]{muskhelishvili}. The
integral we are interested in is
\begin{equation}
\chi'(x) =  \int_a^b \frac{\ud y}{(y-x)^2} \Phi(y)\ .
\end{equation}
This allows for slightly milder behaviour of $\Phi(x)$ near the
end-points:
\begin{equation}
  \Phi(x) \sim {(x-c)^{\beta_c}}\ ,\quad\text{with}\quad 0 \leq \beta_c < 1\ ,
\end{equation}
where we have restricted to real $\beta_c$ for
simplicity. Differentiating the above results w.r.t.~$x$ once, we get
the result
\begin{equation}
  \chi'(x) \sim \mp \pi\beta_c \cot \pi\beta_c\, {(x-c)^{\beta_c-1}}\ .
\end{equation}
Thus, the boundary point analysis yields the following singular terms near $x = 0$ for the right hand side of \eqref{ginvphieq}:
\begin{equation}
  \left(m^2 - \frac{\lambda}{\pi}\right) x^{\beta_0 - 1} + \frac{\lambda}{\pi} \pi\beta_0 \cot \pi\beta_0 x^{\beta_0-1}\ .
\end{equation}
The singular terms vanish if we choose $\beta_0$ such that
\begin{equation}
  \frac{\pi m^2}{\lambda} + \pi\beta_0 \cot \pi\beta_0 = 1\ .
\end{equation}
A similar analysis near $x = 1$ gives the singular term with
$(x-1)^{\beta_1 -1}$
\begin{equation}
-  \frac{\pi m^2}{\lambda} - \pi\beta_1 \cot \pi\beta_1 = -1\ .
\end{equation}

\section{Dimensional regularization of the Euler-Heisenberg action}\label{reg}

Recall the Euler-Heisenberg Lagrangian:
\begin{equation}\label{lagconst}
-\frac{1}{2} \int_0^\infty \frac{\ud t}{(2\pi t)^{d/2}}\e^{-m^2 t} \eta \coth t\eta\ ,
\end{equation}
The above expression is divergent from the $t \sim 0$ (ultraviolet)
region of the integrand and has to be regularized. Following Dittrich
\cite{Dittrich:1975au}, we use the following formula from Gradshteyn
and Ryzhik \cite[3.551, no.~3]{Gradshteyn}:
\begin{equation}
  \int_0^\infty \ud x\, x^{\mu-1} \e^{-\beta x} \coth x = \Gamma(\mu)\left[2^{1-\mu} \zeta(\mu, \beta/2) - \beta^{-\mu}\right]\ ,
\end{equation}
which is valid for $\text{Re}\,\mu > 1$ and $\text{Re}\,\beta > 0$,
where $\zeta(\mu,\beta/2)$ is the Hurwitz $\zeta$-function. The
Lagrangian \eqref{lagconst} can be cast in this form with
$\beta = m^2 / \eta$ and $\mu = 1 - \frac{d}{2}$ by setting
$x =t \eta$:
\begin{align}\label{detevalx}
  &  -\frac{1}{2} \int_0^\infty \frac{\ud t}{(2\pi t)^{d/2}}\e^{-m^2 t}\, \eta \coth t\eta = -\frac{1}{2}\left(\frac{\eta}{2\pi}\right)^{d/2} \int_0^\infty \frac{\ud x}{x^{d/2}} \e^{-m^2 x/\eta}\, \coth x\ ,\nonumber\\
  &\qquad\qquad= -\frac{1}{2}\left(\frac{\eta}{2\pi}\right)^{d/2} \Gamma\left(1-\frac{d}{2}\right) \left[2^{d/2} \zeta\left(1-\frac{d}{2},\frac{m^2}{2\eta}\right) - \left(\frac{m^2}{\eta}\right)^{\frac{d}{2}-1}\right]\ .
\end{align}
We are interested in $d \to 2$. In that case, the $\Gamma$-function on
the right hand side has a pole. Explicitly, writing
$\epsilon = \frac{2-d}{2}$, we have
\begin{equation}
  \Gamma\left(1-\frac{d}{2}\right) = \Gamma(\epsilon) \approx \frac{1}{\epsilon}\ .
\end{equation}
We can Taylor-expand the remaining factors in $\epsilon$ to get the
behaviour near the pole:
% \begin{align}
%   &\zeta\left(1- \frac{d}{2},\frac{m^2}{2\eta}\right) =   \zeta\left(0,\frac{m^2}{2\eta}\right) + \epsilon\, \zeta'\left(0,\frac{m^2}{2\eta}\right) + \mc{O}(\epsilon^2)\ ,\nonumber\\
%   &\left(\frac{\eta}{2\pi}\right)^{d/2} =  \frac{\eta}{2\pi} \left(1 - \epsilon \log \frac{\eta}{2\pi}\right) + \mc{O}(\epsilon^2)\nonumber\\
%   &2^{d/2} = 2(1 -  \epsilon \log 2) + \mc{O}(\epsilon^2)\ ,\quad \beta^{\frac{d}{2}-1} = 1 - \epsilon \log\beta + \mc{O}(\epsilon^2)\ .
% \end{align}
% Thus, the pole in $\epsilon$ is
\begin{align}
  &\frac{1}{\epsilon} \times -\frac{1}{2} \frac{\eta}{2\pi}\left[2 \zeta\left(0,\frac{m^2}{2\eta}\right) - 1\right] = -\frac{\eta}{4\pi\epsilon} \left[1 - \frac{m^2}{\eta} - 1\right] = \frac{m^2}{4\pi \epsilon}\ ,
\end{align}
where, in going to the second equality, we have used
\begin{equation}
  \zeta(n,x) = -\frac{1}{n+1} B_{n+1}(x)\ ,\quad B_1(x) = x - \frac{1}{2} \ ,
\end{equation}
with $B_n(x)$ being the $n^{\rm th}$ Bernoulli polynomial. We can
rewrite the pole as
\begin{equation}
  \frac{m^2}{4\pi \epsilon} = -\frac{m^2}{4\pi} \frac{-1}{\epsilon} = -\frac{m^2}{4\pi} \Gamma\left(-\frac{d}{2}\right)\ .
\end{equation}
So, the above pole can be minimally-subtracted by including the extra
terms in the integrand of \eqref{detevalx}:
\begin{equation}
 \coth x\ \to\  \coth x - \frac{1}{x}\ .
\end{equation}
The finite part in \eqref{detevalx} as $\epsilon \to 0$ is
\begin{equation}
 - \frac{\eta}{4\pi}\bigg[ \frac{m^2}{\eta} \gamma_E + 2 \zeta'\left(0,\frac{m^2}{2\eta}\right) - 2\log 2 \times \zeta\left(0,\frac{m^2}{2\eta}\right) + \log \frac{m^2}{\eta} - \log\frac{\eta}{2\pi} \times \left(2 \zeta\left(0,\frac{m^2}{2\eta}\right) -1\right)\bigg]\ .
\end{equation}
Using $\zeta'(0,a) = \log \Gamma(a) - \frac{1}{2} \log 2\pi$, the above becomes
\begin{align}
  % &-\frac{\eta}{4\pi}\bigg[2\zeta'\left(0,\frac{m^2}{2\eta}\right) -\left(1 - \frac{m^2}{\eta}\right)  \log 2  +  \log \frac{m^2}{\eta} + \frac{m^2}{\eta} \log \frac{\eta}{2\pi}\bigg]\ ,\nonumber\\
  % &=-\frac{\eta}{4\pi}\bigg[2 \log \Gamma\left(\frac{m^2}{2\eta}\right) - \log 2\pi -\left(1 - \frac{m^2}{\eta}\right)  \log 2  +  \log \frac{m^2}{\eta} + \frac{m^2}{\eta} \log \frac{\eta}{m^2} + \frac{m^2}{\eta} \log \frac{m^2}{2\pi}\bigg]\ ,\nonumber\\
  &-\frac{\eta}{4\pi}\bigg[2 \log \Gamma\left(\frac{m^2}{2\eta}\right) - \log 2\pi +\left(1 - \frac{m^2}{\eta}\right)  \log  \frac{m^2}{2\eta}\bigg] - \frac{m^2}{4\pi} \left(\log\frac{m^2}{2\pi} + \gamma_E \right)\ .
\end{align}
Thus, the final form of the log determinant is
\begin{multline}\label{logdetdimreg}
-\frac{\i \eta}{2} \int_0^\infty \frac{\ud t}{(2\pi t)^{d/2}}\e^{-m^2 t} \coth t\eta\bigg|_{d=2,\rm dim-reg.} =  -\frac{\i}{4\pi} \int_0^\infty \frac{\ud t}{t^2}\e^{-m^2 t} (t\eta \coth t\eta - 1)\ , \\ = -\frac{\i\eta}{4\pi}\left[2 \log\Gamma\left(\frac{m^2}{2\eta}\right) - \log 2\pi + \left(1 - \frac{m^2}{\eta}\right) \log \frac{m^2}{2\eta}\right] - \frac{\i m^2}{4\pi} \left(\log\frac{m^2}{2\pi} + \gamma_E \right)\ .
\end{multline}

\section{Spectral density for $m\neq 0$}\label{spectral}

 To extract
$\varrho(\mu)$ in our case, we have to take the inverse Laplace transform
of the right hand side. That is,
\begin{equation}\label{invlaprho}
  \varrho(\mu) = m\int_0^\pi \ud\theta\sin\theta \,\frac{1}{2\pi\i} \int_{\gamma-\i\infty}^{\gamma+\i\infty} \ud t\, \e^{t(\mu - m^2)}\, t\eta \coth t\eta\ .
\end{equation}
where $\gamma$ is a positive real number, perhaps infinitesimally
close to zero. Recall that $\eta$ is a function of $\theta$. We can
close the contour by a large semicircle in the left-half plane
$\text{Re}(t) < \gamma$, provided the integrand is suppressed as the
radius of the semicircle goes to infinity. In the left-half plane, we
have
\begin{equation}
  t = \gamma + R (\cos\varphi + \i \sin\varphi)\ ,\quad \varphi \in \left[\frac{\pi}{2}, \frac{3\pi}{2}\right]\ .
\end{equation}
The  $t\coth t\eta$ factor is
\begin{equation}
  \coth \big(\gamma\eta + R\eta(\cos\varphi + \i \sin\varphi)\big)\ ,
\end{equation}
which is bounded as $R \to \infty$. The remaining factors
$\e^{t(\mu - m^2)}t$ become
\begin{equation}
  \e^{\gamma(\mu-m^2)}  \e^{R\cos\varphi(\mu-m^2)} \e^{\i R \sin\varphi (\mu-m^2)}  R \eta (\cos\varphi + \i\sin\varphi)\ .
\end{equation}
The exponential factor $\e^{R\cos\varphi(\mu-m^2)}$ decays as
$R\to\infty$ provided $\mu > m^2$ since $\cos\varphi < 0$. Thus, for
$\mu > m^2$, the contour in the integral \eqref{invlaprho} can be
replaced by a closed contour that closes in the left-half plane. The
poles of the integrand are the points
\begin{equation}
  \text{Poles}:\quad t = \frac{\i n\pi}{\eta}\ ,\quad \text{for all non-zero integer $n$}\ ,
\end{equation}
and the corresponding residues are
\begin{equation}
  \frac{\i n\pi}{\eta} \e^{\i n\pi(\mu-m^2)/\eta}\ .
\end{equation}
Thus, the result of the contour integral is
\begin{align}
  \varrho(\mu) &= m\int_0^\pi\ud\theta\sin\theta\,\sum_{n=-\infty}^\infty \frac{\i n \pi}{\eta} \exp\left(\frac{\i n\pi(\mu-m^2)}{\eta}\right) \ ,\nonumber\\
            &= \int_0^\pi\ud\theta\sin\theta\, \sum_{n=1}^\infty \frac{-2 n\pi}{\eta} \sin\left(\frac{n\pi(\mu-m^2)}{\eta}\right)\ ,\quad \text{for}\quad \mu > m^2\ .
\end{align}
For $\mu < m^2$, the contour can be closed in the right-half plane
$\text{Re}(\gamma) > 0$. The integrand has no poles in the right-half
plane and hence the result of the integral is zero. Thus, we have
\begin{equation}\label{rhosol}
    \varrho(\mu) = m\, \Theta(\mu - m^2) \int_0^\pi \ud\theta\sin\theta\,\sum_{n=1}^\infty  \frac{-2 n\pi}{\eta}\sin\left(\frac{n\pi(\mu-m^2)}{\eta}\right)\ .
\end{equation}
Observe that the sum over $n$ can be written as the derivative of the
periodic $\delta$-function:
\begin{align}\label{rhosol1}
  \varrho(\mu) &=  m\, \Theta(\mu - m^2) \int_0^\pi \ud\theta\sin\theta\,\frac{\partial}{\partial \mu} \sum_{n=1}^\infty  2\cos \left(\frac{n \pi (\mu - m^2)}{\eta}\right)\ ,\nonumber\\
  &= 2\pi m\, \Theta(\mu - m^2) \delta'_P\left(\pi\frac{\mu - m^2}{\eta}\right)\ ,
\end{align}
where the derivative $'$ is w.r.t.~$\mu$, and $\delta_P$ is the
periodic $\delta$-function
\begin{equation}
  \delta_P(x) = \sum_{k=-\infty}^\infty \delta(x - 2k \pi) = \frac{1}{2\pi}\left(1 + 2\sum_{n=1}^\infty \cos n x\right)\ .
\end{equation}
Thus, the spectral density is
\begin{align}\label{rhosol2}
  \varrho(\mu) &=  2\pi m\, \Theta(\mu - m^2) \int_0^\pi \ud\theta\sin\theta \sum_{k=-\infty}^\infty \delta'\left(\pi\frac{\mu - m^2 - 2k \eta}{\eta}\right) \ .
\end{align}
If we assume that $\eta$ is positive for all $\theta$, we can scale it
out of the $\delta$-function:
\begin{align}\label{rhosol3}
  \varrho(\mu) &=  2 m\, \Theta(\mu - m^2) \int_0^\pi \ud\theta\sin\theta\, \eta(\theta) \sum_{k=-\infty}^\infty \delta'\left(\mu - m^2 - 2k \eta(\theta)\right)\ ,\nonumber\\
            &=  2 m\,  \int_0^\pi \ud\theta\sin\theta\, \eta(\theta) \sum_{k=0}^\infty \delta'\left(\mu - m^2 - 2k \eta(\theta)\right)\ .
\end{align}

%\setlength\bibitemsep{0.5\baselineskip}
%\printbibliography
\bibliography{refs}\bibliographystyle{JHEP}

\end{document}